\documentclass[conference]{IEEEtran}
\ifCLASSINFOpdf
\else
\fi
\usepackage{fancyhdr}
\usepackage{hyperref}

\usepackage{algorithm}
\usepackage{algorithmic}
\usepackage{xspace}
\usepackage[font=small,labelfont=bf]{caption}
\usepackage[caption=false]{subfig}
\usepackage{amsmath}
\usepackage{mathtools}
\usepackage{xcolor}
\usepackage{amsfonts}
\usepackage{amssymb}
\usepackage{pifont}
\usepackage{color}

\xspaceaddexceptions{]\}}
\newcommand{\ignore}[1]{}
\newcommand{\mcdls}[0]{\texttt{MC-DLA}\xspace}
\newcommand{\dcdls}[0]{\texttt{DC-DLA}\xspace}
\newcommand{\hcdls}[0]{\texttt{HC-DLA}\xspace}

\newcommand{\mnode}[1]{\texttt{M$_{#1}$}}
\newcommand{\dnode}[1]{\texttt{D$_{#1}$}}

\newcommand{\mcdlsS}[0]{\texttt{MC-DLA(S)}\xspace}
\newcommand{\mcdlsL}[0]{\texttt{MC-DLA(L)}\xspace}
\newcommand{\mcdlsB}[0]{\texttt{MC-DLA(B)}\xspace}

\newcommand{\dls}[0]{\texttt{DLA}\xspace}
\newcommand{\devicelocal}[0]{\texttt{device$_{local}$}\xspace}
\newcommand{\deviceremote}[0]{\texttt{device$_{remote}$}\xspace}

\newcommand{\old}[1]{}

\newcommand{\fig}[1]{Figure~\ref{#1}}
\newcommand{\sect}[1]{Section~\ref{#1}}
\newcommand{\tab}[1]{Table~\ref{#1}}

\newcommand{\featureIn}[0]{\texttt{X}\xspace}
\newcommand{\featureOut}[0]{\texttt{Y}\xspace}
\newcommand{\gradientIn}[0]{\texttt{dY}\xspace}
\newcommand{\gradientOut}[0]{\texttt{dX}\xspace}
\newcommand{\gradientW}[0]{\texttt{dW}\xspace}

\newcommand{\weight}[0]{\texttt{W}\xspace}

\newcommand\blfootnote[1]{%
\begingroup
\renewcommand\thefootnote{}\footnote{#1}%
\addtocounter{footnote}{-1}%
\endgroup
}

\renewcommand{\baselinestretch}{.999}
\title{Beyond the Memory Wall: A Case for Memory-centric HPC System for Deep Learning}

\begin{document}

\author{

\IEEEauthorblockN{
Youngeun Kwon\hspace{2em}Minsoo Rhu}
\IEEEauthorblockA{
KAIST\\
\texttt{\{yekwon, mrhu\}@kaist.ac.kr}\\
}
}


\maketitle
\pagestyle{plain}

%%%%%% -- PAPER CONTENT STARTS-- %%%%%%%%
\begin{abstract} 

As the models and the datasets to train deep learning (DL) models scale, system
	architects are faced with new challenges, one of which is the memory capacity
	bottleneck, where the limited physical memory inside the accelerator device
	constrains the algorithm that can be studied. We propose a memory-centric
	deep learning system that can transparently expand the memory capacity
	available to the accelerators while also providing fast inter-device
	communication for parallel training. Our proposal aggregates a pool of memory
	modules locally within the device-side interconnect, which are decoupled from
	the host interface and function as a vehicle for transparent memory capacity
	expansion.  Compared to conventional systems, our proposal achieves an
	average $2.8\times$ speedup on eight DL applications and increases the
	system-wide memory capacity to tens of TBs.

\end{abstract}

\IEEEpeerreviewmaketitle

% footnote for author version
\blfootnote{
Published as a conference paper at the $51^{\text{st}}$ IEEE/ACM International Symposium on Microarchitecture (MICRO-51), 2018.
}

\section{Introduction}		
\label{sect:intro}

\begin{figure}[t!] \centering
\includegraphics[width=0.43\textwidth]{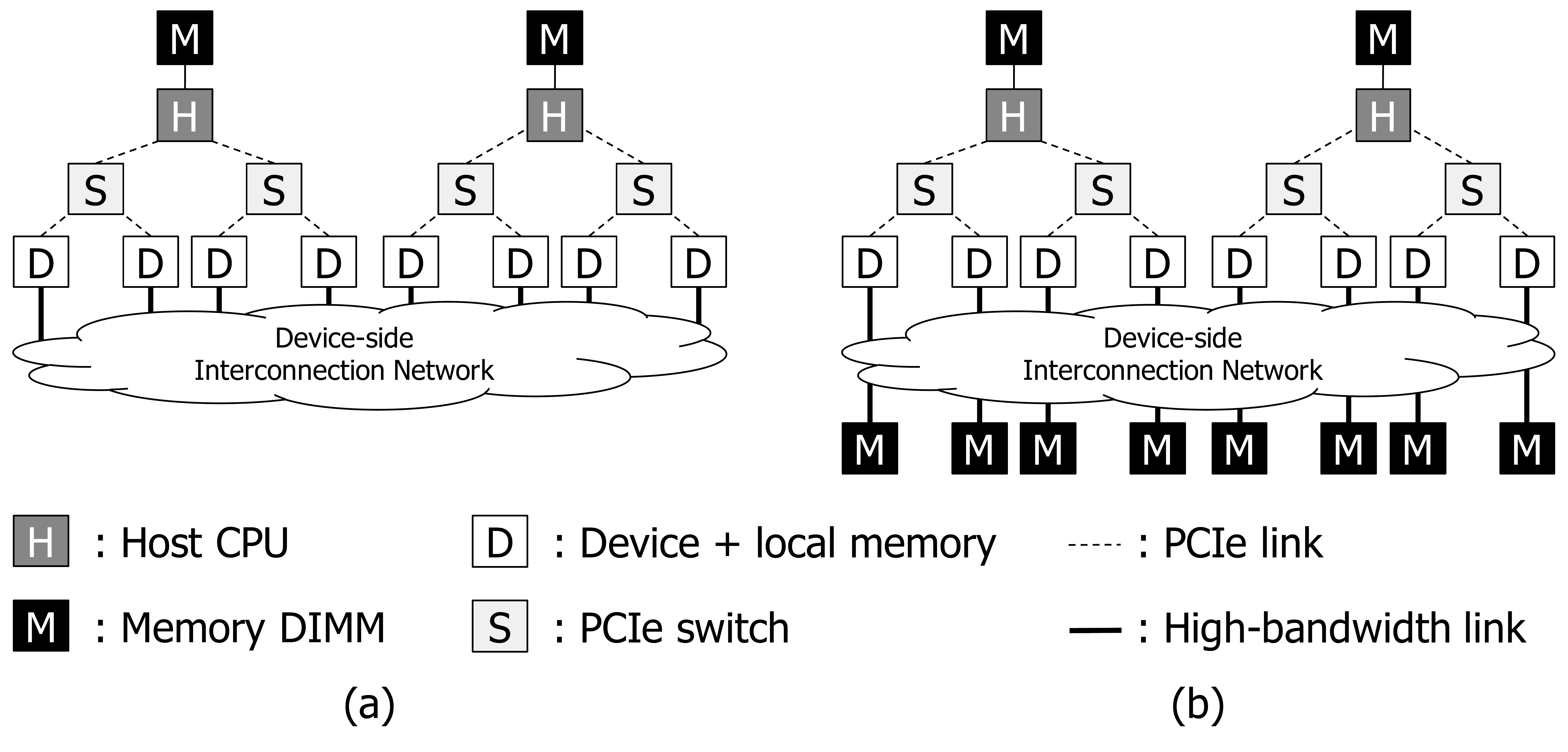}
\caption{
(a) A device-centric deep learning system architecture, and (b) a memory-centric deep learning system architecture.
}
\vspace{-1.2em}
\label{fig:dcdls_vs_mcdls}
\end{figure}

Deep learning (DL) models and its training datasets are scaling at a phenomenal
rate, so the progress in DL is primarily limited by how fast the deep neural
network (DNN) model can be evaluated and how large of a memory you can utilize
for training. DL practitioners are therefore seeking efficient \emph{parallel}
training solutions, increasingly adopting a \emph{dense} system node design,
	 which houses several PCIe-attached co-processor devices~\cite{volta_v100,tpu2,nervana} to increase the
	 node-level throughput. As the system-level performance is contingent
	 upon how the DL algorithm is parallelized across the devices and
	 how effectively they communicate with each other, leading vendors in this
	 space are employing a custom \emph{device-side interconnection network} that
	 utilizes proprietary high-bandwidth signaling solutions (e.g., NVIDIA's
			 NVLINK which provides $100$s of GB/sec of bandwidth) for fast
	 communication and synchronization~\cite{nervana,nvlink}.  Such
	 \emph{device-centric} deep learning system architecture (\dcdls) is becoming
	 mainstream for DL (\fig{fig:dcdls_vs_mcdls}(a)) and we are seeing an
	 increasing number of HPC systems that employ a standalone, device-side
	 interconnection network for efficient DL
	 parallelization~\cite{nervana,dgx_1v,hgx_1,dgx_2}.  
	 
	As researchers seek to deploy deeper and larger DNN topologies however,
	end-users are faced upon a memory ``capacity'' wall, where the limited
	on-device physical memory (based on on-package 3D stack memory~\cite{hbm} in
	NVIDIA's V100~\cite{volta_v100}, Google's TPUv2~\cite{tpu2},
	and Intel-Nervana's Lake Crest accelerator~\cite{nervana})
	constrains the algorithm that can be
	trained~\cite{rhu:2016:vdnn,ibm_lms,dnn_train,tbd_toronto}. Increasing the 
	capacity of these on-package stacked DRAM however is challenging due to
	wireability of the silicon interposers, chip pinout required to drive the
	added DRAM stacks, and technology limits on how many DRAM stacks you can
	vertically integrate.  Consequently, recent solutions have proposed to use
	the device (GPU/TPU) memory as an application level cache with respect to the
	host CPU
	memory~\cite{rhu:2016:vdnn,ibm_lms,meng:2017:gpu_mem_opt,park:2018:ppopp,vdnn_in_chainer,vdnn_in_tf},
	effectively \emph{virtualizing} DNN memory usage across the host and device
	memory via PCIe (\sect{sect:vdnn}).  The effectiveness of these prior
	solutions however is sensitive to the host-device communication bandwidth as
	it determines the latency incurred in migrating DNN data in/out of these two
	memory regions.  The left-axis in \fig{fig:motivation} shows the execution
	time of state-of-the-art convolutional neural networks (CNNs) on successive
	versions of a \emph{single}, high-end DL accelerator, which has been reduced
	by a factor of $20\times$ to $34\times$ over five years. During these time
	periods, the signaling circuitry and the overall communication bandwidth
	offered by the latest PCIe and InfiniBand has improved only by a factor of
	$1\times$ (PCIe gen3) and $3.5\times$ (i.e., IB-FDR to IB-HDR), respectively.
	This has led to a steadily increasing performance overhead of host-device
	memory virtualization via PCIe (right-axis in \fig{fig:motivation}),
	where the growing performance gap between the device computing power and
	host-device (PCIe) communication bandwidth aggravates system-level
	performance. Virtualizing DNN memory over a \emph{multi}-GPU/TPU system
	incurs even higher performance overheads because the effective host--device
	communication bandwidth allocated per device gets proportionally reduced to
	the number of intra-node devices. Therefore, the overall system can
	experience a significant performance slowdown due to the additional latency
	incurred during host--device memory copies (\sect{sect:results}).  Overall,
	current trends point to an urgent need for a system architectural solution that
	satisfies the dual requirements of (a) fast inter-device communication for
	parallel training, and (b) high-performance memory virtualization over a
	large memory pool to enable memory hungry DNNs to be trainable over accelerator
	devices.

		In this paper, we make a case for a \emph{memory-centric} deep learning
		system architecture (\mcdls) that aggregates a pool of capacity-optimized
		memory modules within the device-side interconnect for
		transparent memory capacity expansion (\fig{fig:dcdls_vs_mcdls}(b)).  
While our proposal is
		reminiscent of prior disaggregated memory proposals~\cite{disagg_mem_1,disagg_mem_2}, the
		CPU-centric memory disaggregation solutions suffer from the same
		performance bottlenecks of \dcdls because of its reliance on PCIe. In our proposal,
		the pool of memory modules (henceforth referred to as \emph{memory-nodes}) 
		are completely decoupled from the	legacy, host-device interface (e.g., PCIe) and
		are stationed \emph{locally} within the device-side interconnect. We propose
		to interconnect the accelerators and the memory-nodes using 
		the high-bandwidth, low-latency signaling links (e.g.,
				NVLINK)	and utilize the memory-nodes as a backing store
		to the accelerators. This allows the memory-nodes to function as a vehicle
		for transparent memory capacity expansion, allowing researchers to
		train DL algorithms that are much larger, deeper, and more complex. Because the
		accelerators access the memory-nodes via the high-bandwidth links, the
		performance overhead of virtualizing memory can be substantially
		reduced.  At the same time, \mcdls connects the accelerators and the
		memory-nodes in a manner that maximizes inter-device communication
		bandwidth so that \dcdls's high-efficiency in conducting collective
		communication operations is maintained.  Overall,
		this paper makes the following contributions:

\begin{itemize} 

\item To the best of our knowledge, our work is the first that highlights the
importance of device-side interconnects in training scaled up DL algorithms,
presenting a quantitative analysis on parallel training in the context of HPC systems with multiple accelerator (GPU/TPU) devices.

\item This work identifies key system-level performance
bottlenecks on \dcdls and motivates the need for a new 
system architecture that balances fast communication and user productivity in training large DNN algorithms.

\item We propose and evaluate a system architecture called \mcdls that
provides transparent memory capacity expansion while also 
enabling fast inter-device communication. Compared to \dcdls designs, our proposal
achieves an average $2.8\times$ performance improvement while expanding
the system-wide memory capacity exposed to the accelerators to tens of TBs.

\end{itemize}

\section{Background and Motivation}
\label{sect:motivation}

\subsection{DL Training versus Inference}
\label{sect:training}

\begin{figure}[t!] \centering
\includegraphics[width=0.44\textwidth]{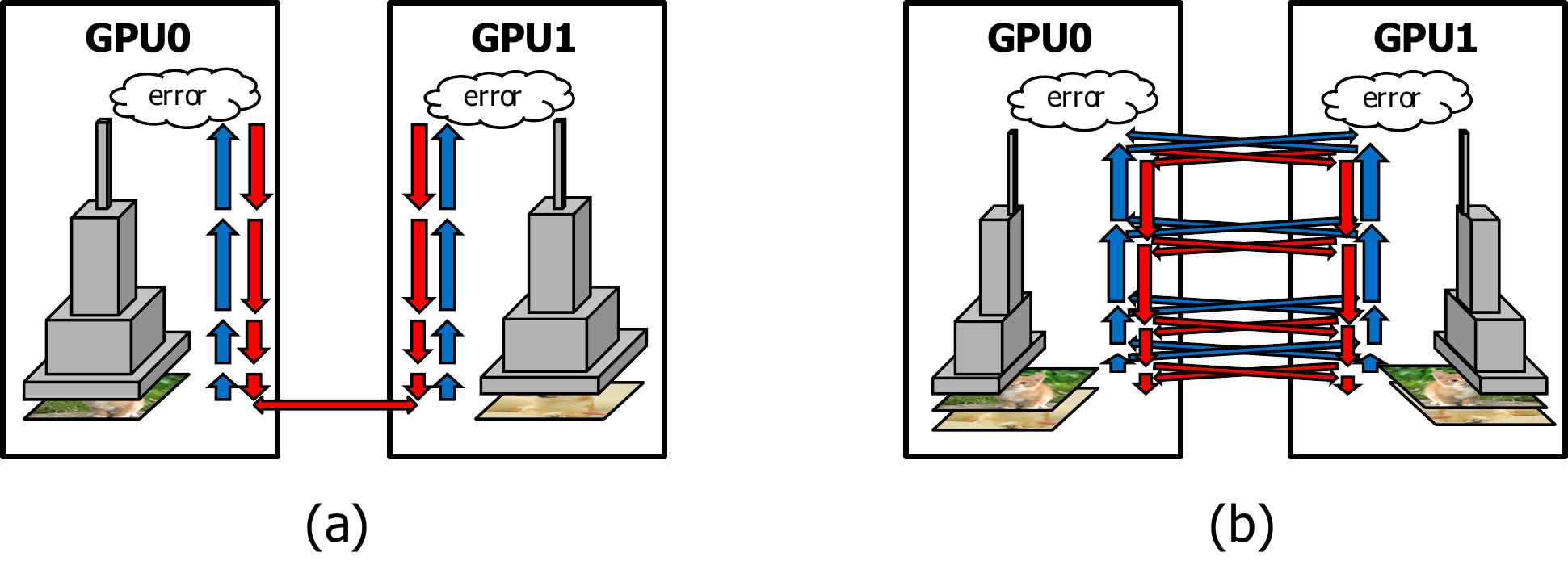}
\caption{ (a) Data-parallel and (b) model-parallel training. Blue and red
	arrows inside each device represent a given layer's computation during
		forward and backward propagation, respectively. The blue and red
		arrows that crosses the boundaries of two devices represent the
		per layer inter-device communication and synchronization operations. As shown,
		model-parallel training incurs much more frequent synchronization than
		data-parallel training~\cite{dgx_1_white_paper,alex_weird_trick}.  
}
\vspace{-1.2em}
\label{fig:parallel_dl_training}
\end{figure}

DNNs require \emph{training} to be ready for
\emph{inference}.  Training is a three-step process that involves learning the
optimal values of the DNN weights using the backpropagation
algorithm~\cite{lecun_gd}.  First, a serialized, layer-wise computation process
called \emph{forward propagation} is taken from the first (input) layer to the
last (output) layer in a serialized fashion (the blue arrows from bottom to top
		in \fig{fig:parallel_dl_training}). A given layer applies a set of
mathematical operation (e.g., convolution, activation, recurrence, etc) to the
input \emph{feature maps} (\featureIn) and derives the output feature maps
(\featureOut), which is forwarded to the next layer to be used as its input
feature maps.  At the end of forward propagation, a prediction
of the input is given, which is compared against the ground truth. The defined
loss function quantifies the magnitude of the error between the current
prediction and the ground truth, which is encapsulated in a value called
\emph{gradients} of the loss function with respect to the last layer's output.
Then, \emph{backpropagation} is performed in the opposite direction of forward propagation
(the red arrows from top to bottom) again in a layer-wise manner, where the
incoming gradients (\gradientIn) are used to derive the output gradients
(\gradientOut) to be sent to the previous layer to be used as its input
gradients.  Using the \gradientIn and \gradientOut, each layer derives its own
weight gradients (\gradientW) to adjust its own layer's weights (\weight) so that the
loss value is incrementally reduced, improving the performance of the DNN model.

\begin{figure}[t!] \centering
\includegraphics[width=0.49\textwidth]{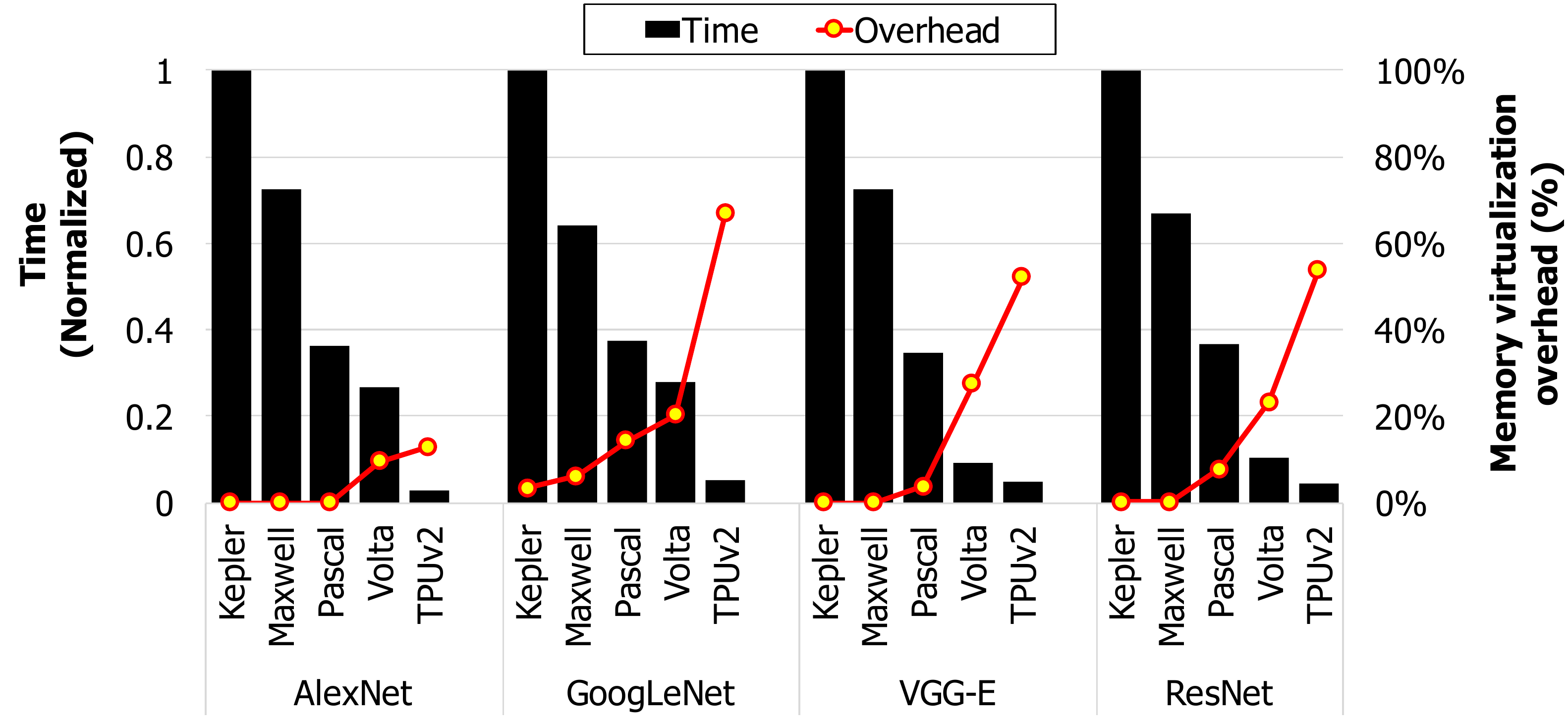}
\caption{
Execution time of 
	running state-of-the-art CNN models across five recent generation of a single DL accelerator device (left-axis) 
	and the performance overhead incurred due to memory
	virtualization (right-axis). See \sect{sect:eval} for evaluation methodology.
}
\vspace{-1.2em}
\label{fig:motivation}
\end{figure}

\subsection{Virtualizing Memory for Deep Learning}
\label{sect:vdnn}
The chain-rule based backpropagation algorithm requires a given layer's input
feature maps values (\featureIn) to derive the gradient values of the layer's
weights (\gradientW)~\cite{lecun_gd}.  Consequently, the overall memory
allocation size of DNN training scales proportionally to the network depth
(i.e., memory cost of \emph{O(N)} to train a DNN with \emph{N} layers).
End-users must therefore carefully tune their network topology (i.e., the
		number of layers in a DNN and the inter-layer connections) and the training
batch size to make sure the overall memory requirement fits within the physical
memory capacity, which can severely limit user productivity. Given recent
research trends where DL practitioners are seeking to deploy ever larger and
deeper network algorithms (e.g., the memory allocations required for training
		can easily exceed $100$s of
		GBs~\cite{resnet,nn_stochastic_depth,densenet,deepbench,splitnet,seq2seq_video_to_text,read_write_mem_network}),
			 tackling this memory capacity bottleneck while minimizing performance
			 overheads becomes vital in enabling researchers to keep studying scaled
			 up DL algorithms. Prior work on virtualizing memory usage of
			 DNNs~\cite{rhu:2016:vdnn,ibm_lms,meng:2017:gpu_mem_opt,park:2018:ppopp,vdnn_in_chainer,vdnn_in_tf}
			 have proposed to utilize both host and device memory concurrently for
			 allocating data structures for DNN training.  By leveraging the
			 user-level DNN topology graph as means to extract a compile-time data
			 dependency information (which is encapsulated as a direct acyclic graph
					 (DAG) data structure)	of the memory-hungry data
			 structures, e.g., feature maps (\featureIn) and/or weights (\weight),
			 DNN virtual memory can leverage this data dependency information to
			 derive the DNN data \emph{reuse distance} to schedule performance-aware
			 data copy operations via \emph{memory-overlaying} across host and device
			 memory via PCIe~\cite{gpu_paging,gpu_x86_at,gpu_tlb}. %Note that host-device memory virtualization via

			 Existing DL frameworks~\cite{ibm_lms,vdnn_in_chainer,ibm_ddl} therefore
			 opt to leverage this DAG to schedule software-level
			 DMA initiated data transfers between host and device, overlapping it with the DNN
			 forward and backward propagation, to maximally utilize the PCIe
			 communication bandwidth and minimize the performance overheads of data
			 migration. By only keeping soon-to-be used DNN data inside the device
			 memory, the memory allocation size of training a network with $N$ layers
			 can be reduced from \emph{O(N)} to \emph{$O(1)$}, enhancing DL
			 practitioners' ability to train scaled-up algorithms.

\subsection{Parallelization of DL Algorithms}
\label{sect:parallel_dl_training}

As the DNN algorithm gets more complex and
deeper~\cite{resnet,nn_stochastic_depth,densenet}, the need for
distributed multi-node systems, each with multiple accelerator devices, have
significantly increased to provide high computing horsepower for DL
practitioners.  Consequently, efficient parallelization of DL algorithms and
fast communication among the devices become vital for maximally exploiting the
HPC systems based on these dense multi-device nodes. Note that the scope of
this paper is on developing an efficient \emph{intra}-node system architecture,
		 so as in conventional designs, we assume that \emph{inter}-node
		 communication is handled using MPI via Ethernet or InfiniBand.

{\bf Parallel DL Training.} The most popular parallel training
strategies employed by DL frameworks are \emph{data-parallel} and
\emph{model-parallel} training (\fig{fig:parallel_dl_training}). 
		Data-parallel training is a
		parallelization scheme that allocates the same network model across all
		the workers, but each worker is assigned with a different
		batch of the overall training dataset. In
		model-parallel training however, all workers work on an identical batch of the
		training dataset (i.e., the problem size is fixed at batch size
			\texttt{N}), but each are allocated with different portions of the
		network model. The parallel tasks
		distributed across the workers must periodically \emph{synchronize} to
		have a consistent DNN model trained within each worker, preventing both
		data-parallel and model-parallel training from achieving perfect scaling.
		Model-parallel training generally incurs much frequent synchronization
		than data-parallel approaches as the input feature maps (\featureIn)
		and gradients (\gradientOut, \gradientW) must be aggregated across layer boundaries 
		due to the nature of its parallelization algorithm. Data-parallel training, on the other hand,
		only requires the accumulation of \gradientW during backpropagation and is therefore
		assumed to be more amenable for achieving close to linear speedup. However, not all networks or 
		layers can be data-parallelized easily, especially for DNNs with large models~\cite{dean:2017:nips,splitnet}, so both
		model and data-parallel training are considered important in quantifying the
		robustness of the system interconnect design of HPC systems.

 \begin{figure}[t!] \centering
\includegraphics[width=0.40\textwidth]{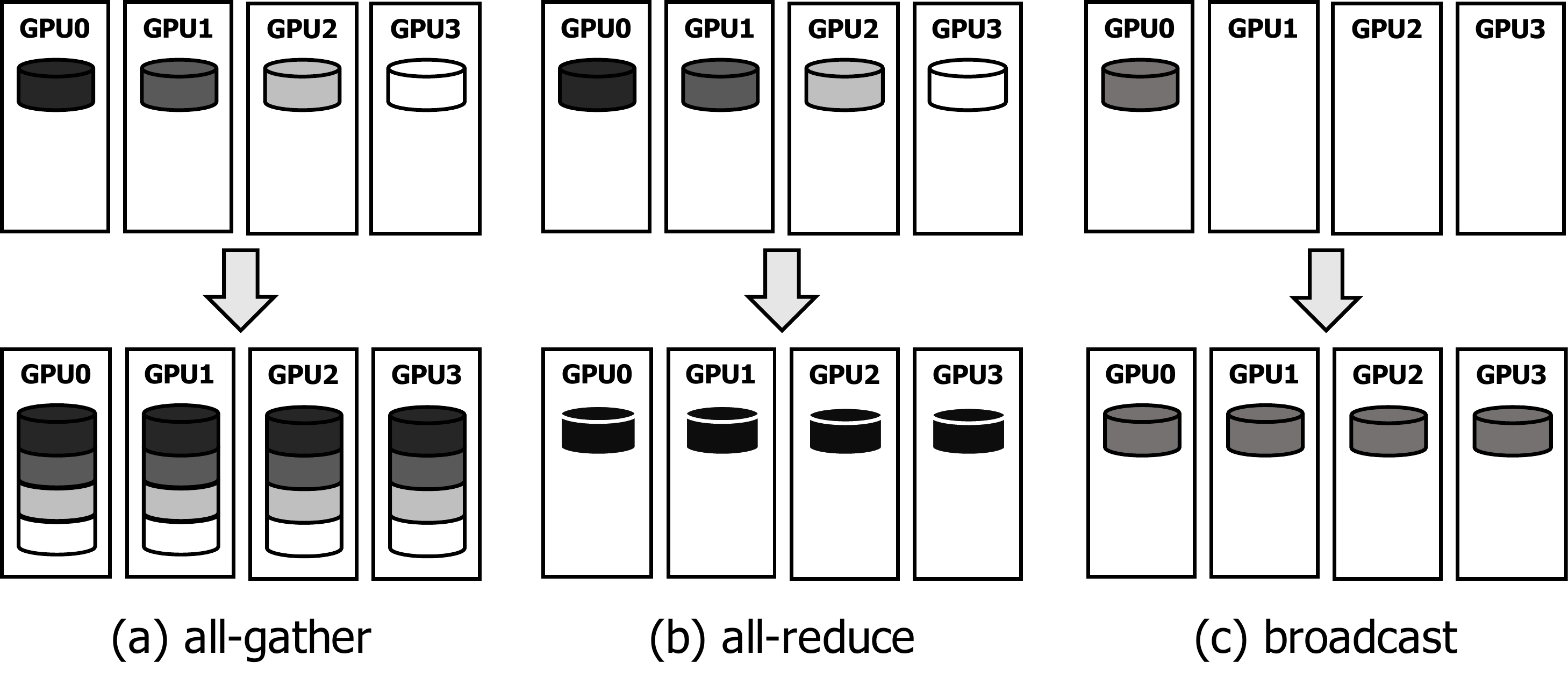} 
\caption{
Key collective communication primitives for parallel training.
}
\vspace{-1.2em}
\label{fig:collective_comm}
\end{figure}

{\bf Communication.} As implied through previous discussions, minimizing
communication overheads is key in high-performance parallel training.
Consequently, maximally utilizing the communication bandwidth provisioned
across accelerators within and across compute nodes is crucial.  Key collective
communication primitives for parallel training are all-gather
(\featureIn), all-reduce (\gradientOut and \gradientW), and broadcast
(\gradientW), which are shown in \fig{fig:collective_comm}. Prior
work~\cite{chan2006collective,nccl:wooley} has demonstrated that the \emph{ring-algorithm}
based collective communication can provide optimal link bandwidth
utilization for the aforementioned collective operations. 
Leading system vendors in this space are therefore employing a topology-aware,
				ring-algorithm based collective communication library (e.g., NVIDIA's NCCL~\cite{nccl}, IBM's PowerAI DDL~\cite{ibm_ddl},
		and Baidu's AllReduce~\cite{baidu_comm}). These libraries cast the underlying system interconnect
as multiple ring networks and orchestrate the DL communication operations
based on the ring-algorithm for maximizing bandwidth utilization while minimizing
latency.

\begin{figure}[t!] \centering
\includegraphics[width=0.27\textwidth]{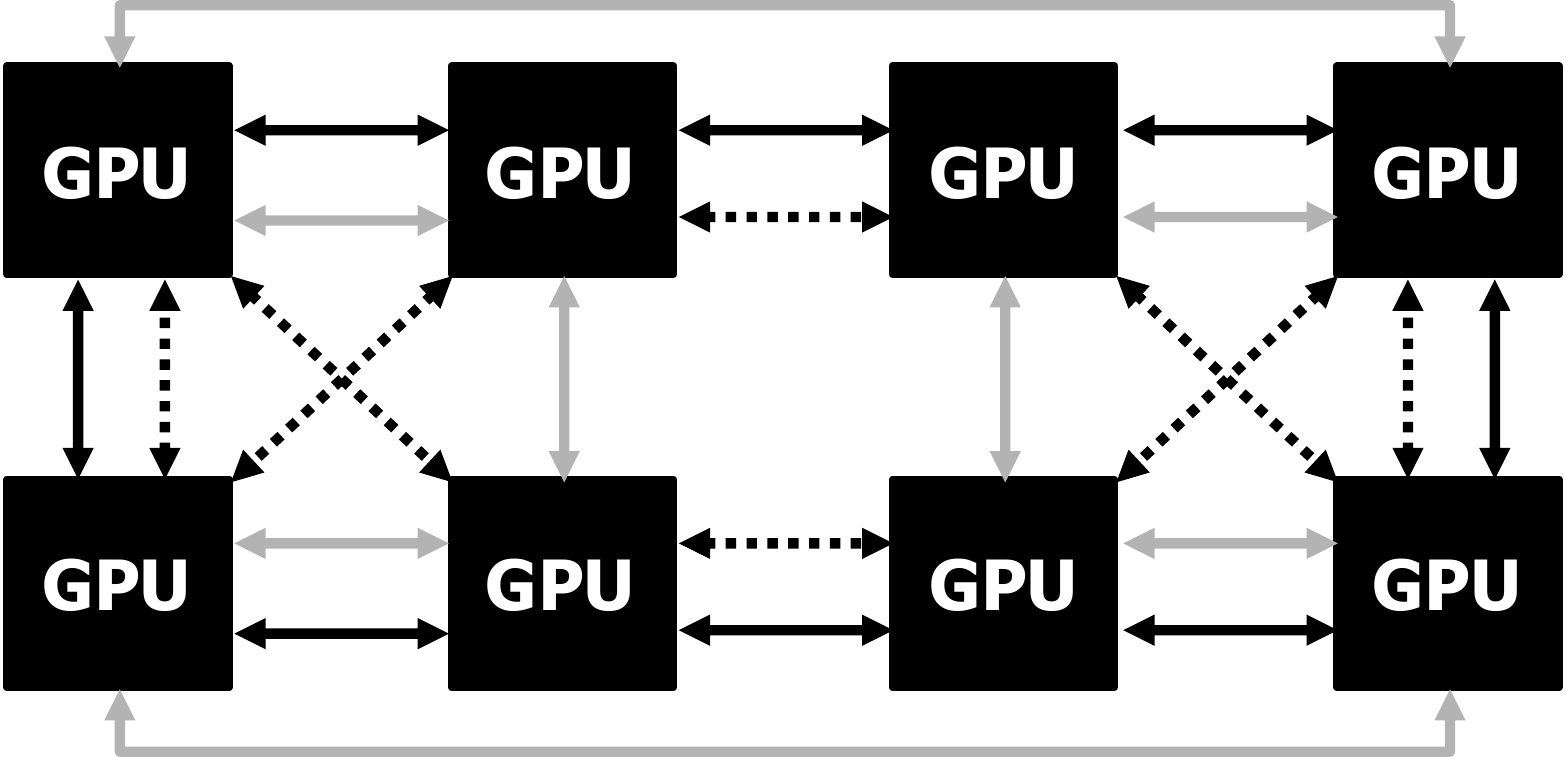}
\vspace{0.5em}
\caption{
The cube-mesh, device-side interconnect employed in NVIDIA's DGX system. The
black, gray, and dotted arrows form three ring networks for collective
communication operations~\cite{dgx_1v,dgx_2}.
}
\vspace{-1.2em}
\label{fig:dgx_1v}
\end{figure}

{\bf Device-side Interconnects for DL.} For efficient communication and
synchronization across accelerator devices, recent HPC systems for DL are
employing proprietary, high-bandwidth \emph{device-side interconnection
	networks} that provide $100$s of GB/sec of inter-device communication
	bandwidth. Intel-Nervana's Lake Crest accelerator~\cite{nervana} employs $12$
	high-bandwidth signaling links ($20\times$ that of what PCIe provides) that
	can tightly couple the DL accelerator devices with each other. NVIDIA's DGX
	system~\cite{dgx_1v} is equipped with $8$ Volta V100~\cite{volta_v100} GPUs
	where each V100 comes with $6$ high-bandwidth NVLINKs (bi-directional
			bandwidth of $50$ GB/sec per link, aggregate channel bandwidth of $300$
			GB/sec per GPU), which are used to form a cube-mesh topology across eight
	V100s (\fig{fig:dgx_1v}). By casting the cube-mesh topology as three ring
	interconnects, the eight GPUs communicate through these high-bandwidth ring
	networks using the NCCL library, which helps achieve optimal bandwidth
	utilization and minimize latency. While such \emph{device-centric} deep
	learning system architecture (\dcdls) solution has advantages in terms of
	inter-device synchronizations, communicating with the host-side CPU can only
	be done using the legacy PCIe link, which can cause severe performance
	bottlenecks for devices utilizing the host CPU memory for memory
	virtualization. 
	
	One might expect that a system architecture that is optimized
	in the other end of the spectrum, where the high bandwidth links are all or
			partially used to access CPU memory, could achieve the best of both
	system-level communication (that is, providing decent communication bandwidth using the
			now singular or duo ring networks) and high-performance virtual memory
	(having high enough bandwidth to read/write to/from CPU memory). Such
	\emph{host-centric} DL system architecture (\hcdls) however falls
	short on several aspects. First, an \hcdls system is simply not a feasible
	option to begin with for the vast majority of the current, x86-based HPC
	systems because these proprietary high-bandwidth signaling links for
	device-side interconnects are incompatible with x86 CPUs. \hcdls systems that
	do enable high-bandwidth CPU-GPU communications (e.g., IBM Power \texttt{+} NVIDIA GPUs)
	still face the following challenges. 
	First, allocating a subset of the high-bandwidth links to connect to the host CPU
	leaves smaller inter-device bandwidth available for device communication,
	which could potentially slowdown the effective node-level throughput for
	algorithms sensitive to inter-device communications. More importantly, however,
	having just a single high-bandwidth link per each accelerator device directly
	connected to the host CPU can leave little memory bandwidth available to the
	CPU itself, leading to a highly unbalanced system design. As we detail in the
	next section, virtualizing the memory usage of DL algorithms require the DMA
	engine to fully utilize the communication link bandwidth in order to
	effectively hide the latency of copying data in and out of device memory.
	This means that a singular high-bandwidth link of $25$ GB/sec per device
	would amount to a total of $100$ GB/sec of worst-case host-side memory
	bandwidth consumption when accounting for the four PCIe-attached devices
	connected to a single CPU socket. As a point of reference, the maximum memory
	bandwidth available for a high-end Intel Xeon CPU and the IBM Power9 is
	``only'' $80$ GB/sec~\cite{intel_xeon} and $120$ GB/sec~\cite{ibm_power9} per
	socket, respectively, due to CPU's latency-oriented design (rather than the
			throughput-oriented GPUs or Google's TPUs~\cite{tpu2} that require high
			bandwidth rather than low latency). As quantitatively discussed in
	\sect{sect:results}, \hcdls can consume an average $92\%$ of host-side memory
	bandwidth for certain workloads, leaving only $8\%$ of memory bandwidth
	available for the host CPU itself. While we explore such design point in this
	paper, we argue that such unbalanced system
	architecture is less practical for future DL systems as it severely
	lacks design flexibility; that is, the amount of read/write throughput the
	system designer can provision for host-device memory virtualization is
	limited by the maximum memory bandwidth available per each CPU socket,
	regardless of how much device-side high-bandwidth links are available, which
	can cause severe bottlenecks for future algorithms that are much larger, deeper,
	and more complex.

\section{Memory-centric HPC System Architecture \\for Deep Learning}
\label{sect:mcdls}

In this paper, we propose a new architectural solution for future HPC systems
optimized for deep learning.  Our goal is to develop a DL system architecture that
		enables \emph{fast inter-device communication for parallel training} while
		at the same time \emph{provisioning high-bandwidth communication channels
			to a pool of capacity-optimized memory modules} (which we refer to as
					\emph{memory-nodes} in the rest of this paper) for high-performance
			virtual memory.  We argue that HPC system architectures for DL training
			should be designed in a \emph{memory-centric} manner as the memory
			``capacity'' wall poses one of the biggest challenges in training deep
			and large learning algorithms~\cite{rhu:2016:vdnn,ibm_lms,dnn_train}.
			Prior work on disaggregated memory~\cite{disagg_mem_1,disagg_mem_2} can
			similarly expand the pool of memory exposed to the system through a
			separate memory-blade accessed over PCIe or the NIC. Similar to \dcdls
			however, the growing performance gap between (GPU/TPU) device computing
			power and host-device communication (\fig{fig:motivation}) renders
			the CPU-centric, PCIe-based memory disaggregation solutions impractical for deep learning
			training as the latency to access the added memory pool will become bottlenecked
			by PCIe.  Consequently, the memory-nodes in our memory-centric DL system
			architecture (\mcdls) are stationed \emph{locally} inside the device-side
			interconnection network, eliminating all its ties with the host PCIe
			interface. This section details the design of the memory-node
			architecture and its application for our proposed \mcdls system that
			leverages these memory-nodes as building blocks to achieve the
			aforementioned design goals. As the scope of this paper is on
			studying the \emph{intra}-node system architecture, 
			we refer to the PCIe-attached accelerator devices (e.g., GPUs
					or TPUs) as \emph{device-nodes} in the rest of this paper because both the memory-nodes and
			device-nodes function as separate nodes inside the device-side
			interconnect (\fig{fig:dcdls_vs_mcdls}(b)).  \mcdls is applicable 
			for both GPUs and TPUs as our proposal concerns
			an efficient \emph{system} architecture design for DL accelerators (i.e., the
					device-nodes). For ease of explanation, we assume the
			device-nodes are based on GPUs and use terminologies defined in
			NVIDIA's CUDA hardware/software interface in the remainder of this
			section.

\subsection{Memory Node Architecture}
\label{sect:m_node}

\begin{figure}[t!] \centering
\includegraphics[width=0.47\textwidth]{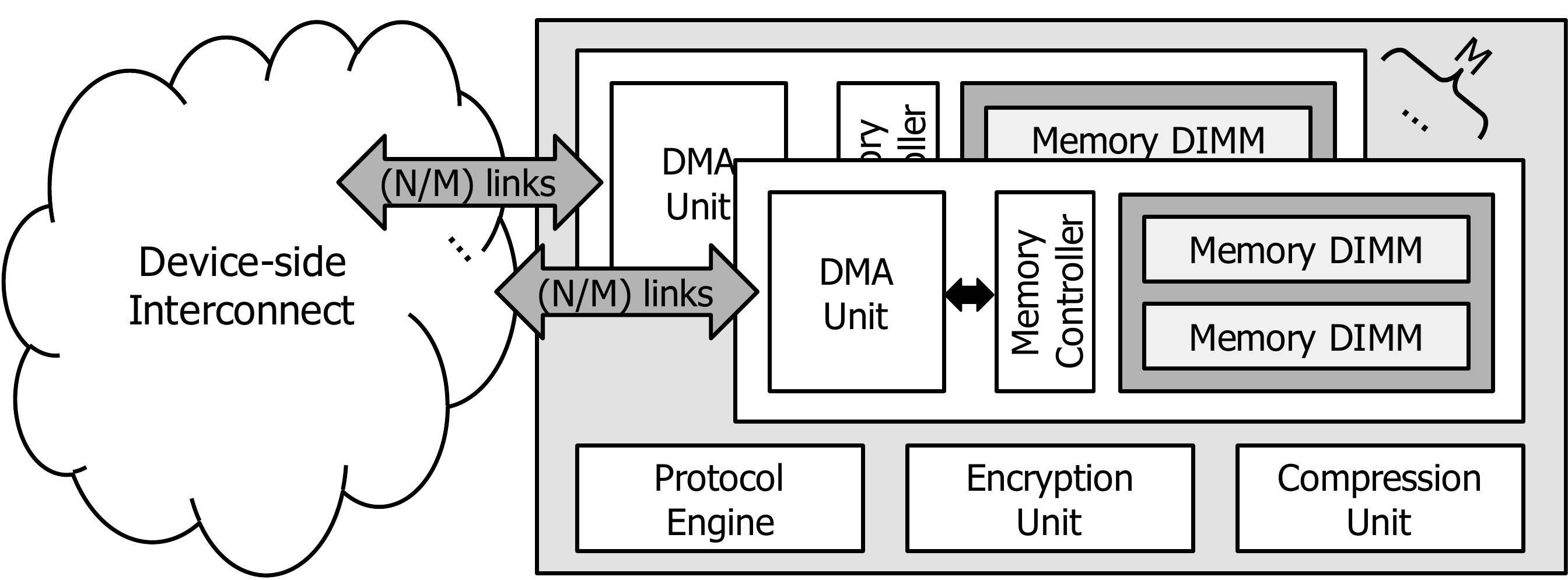}
\caption{Memory-node architecture.}
\vspace{-1.2em}
\label{fig:memory_node}
\end{figure}

The key objective of our memory-node design is to unlock the high-bandwidth
communication channels of the device-side interconnect for high-performance
virtual memory.  \fig{fig:memory_node} illustrates the design of our
memory-node architecture, which contains \texttt{N} high-bandwidth links for
communicating with the device-side interconnection network.  The \texttt{N}
links are logically partitioned into \texttt{M} groups
(\texttt{M}$\leq$\texttt{N}) and each group of (\texttt{N/M}) links are used
exclusively by a designated device-node for DNN memory virtualization.  A
protocol engine that is compatible with the device-side interconnect is used to
provide a maximum bandwidth of \texttt{B} GB/sec per link, so a device-node
assigned with a group of (\texttt{N/M}) links can utilize the DMA engine to
read (write) data from (to) the memory DIMMs with
(\texttt{N/M})$\times$\texttt{B} GB/sec of throughput.  The DMA engine forwards
a device-node's data transfer request to the memory controller which has an
array of commodity memory DIMMs it manages. An ASIC that handles encryption or
compression can optionally be added to the memory-node.  This paper assumes
that the memory DIMMs are populated with capacity and density optimized
commodity memory solutions: from $8$--$16$ GB DDR4 RDIMMs (registered DIMMs) to
$32$--$128$ GB LRDIMMs (load-reduced DIMMs). To narrow down the design space we
explore, the rest of this paper assumes that the board housing a single
memory-node is sized equivalent to a high-end PCIe accelerator board to be
compatible with existing device-side interconnects and minimize the design
costs of the server chassis enclosure. A memory-node built out of a mezzanine
board sized equivalent to Volta V100's ($14$ cm $\times$ $8$ cm) can 
house ten DDR4 DIMMs, providing a maximum of $170$ GB/sec (PC4-$17000$) to
$256$ GB/sec (PC4-$25600$) of memory bandwidth with an overall memory capacity
expansion of $80$ GB to $1.3$ TB per memory-node.

\subsection{System Architecture}
\label{sect:sys_arch}

As our work is the first that highlights the importance of device-side interconnects
in training scaled-up DL algorithms, system architects are given a wide design space
under our proposal. A full design space exploration is beyond the scope of this
paper, so this section presents three system interconnect design points
that incorporate our memory-nodes and discuss their trade-offs in terms of link
bandwidth utilization and overall performance. To narrow down our design options, 
we assume that the number of device-nodes and memory-nodes are identical and
that all device-nodes and memory-nodes have \texttt{N} high-bandwidth
communication links to interface with the other nodes in the network (each
		link providing \texttt{B} GB/sec of uni-directional communication
		bandwidth, \fig{fig:memory_node}).  We use the system
configuration of NVIDIA's DGX system (\texttt{N}=$6$ high-bandwidth links per
		device, each link providing \texttt{B}=$25$ GB/sec communication bandwidth)
as a running example to describe the design intuitions behind \mcdls.

\begin{figure}[t!] 
\centering
\subfloat[]{
	\includegraphics[width=0.34\textwidth]{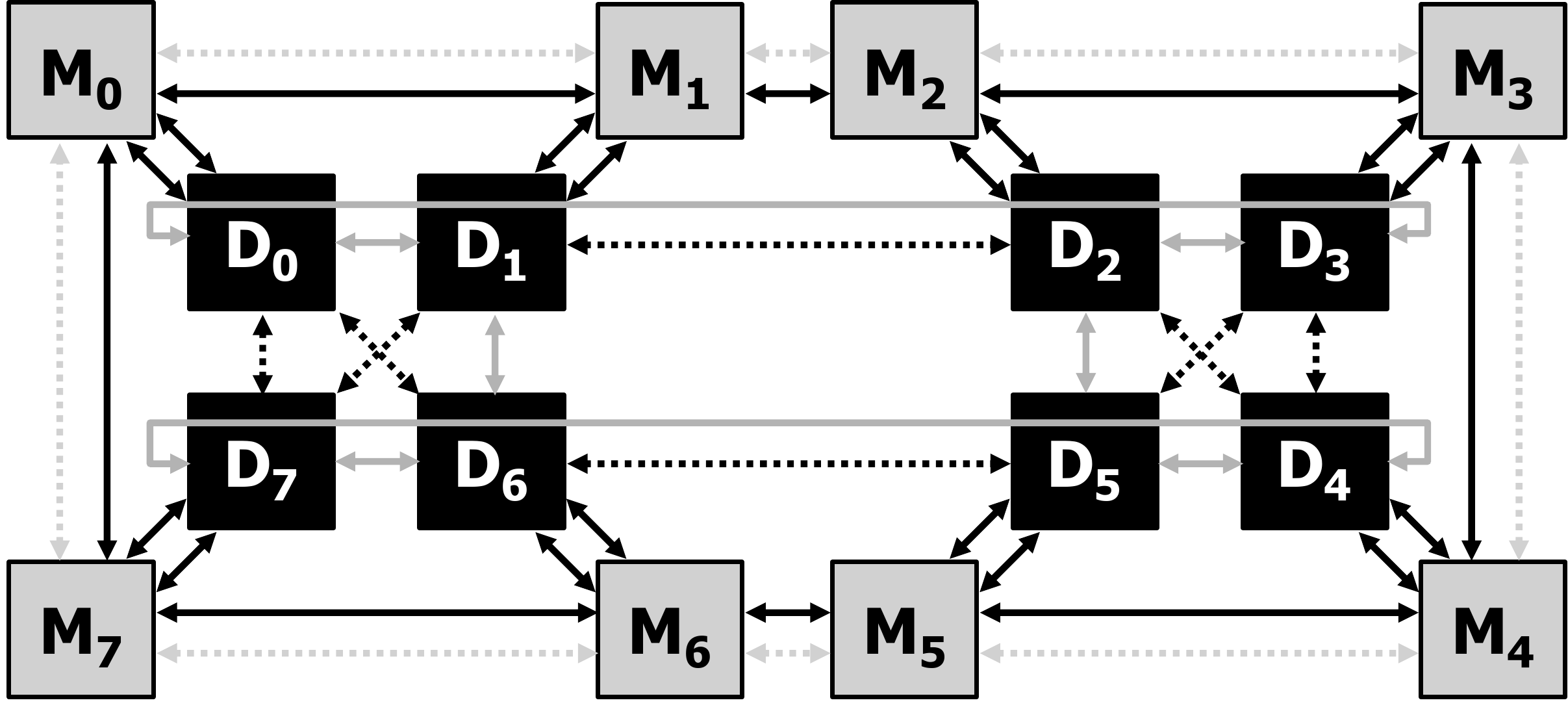}
	\label{fig:mcdla_d_star}
}
\vspace{1em}
\subfloat[]{
	\includegraphics[width=0.34\textwidth]{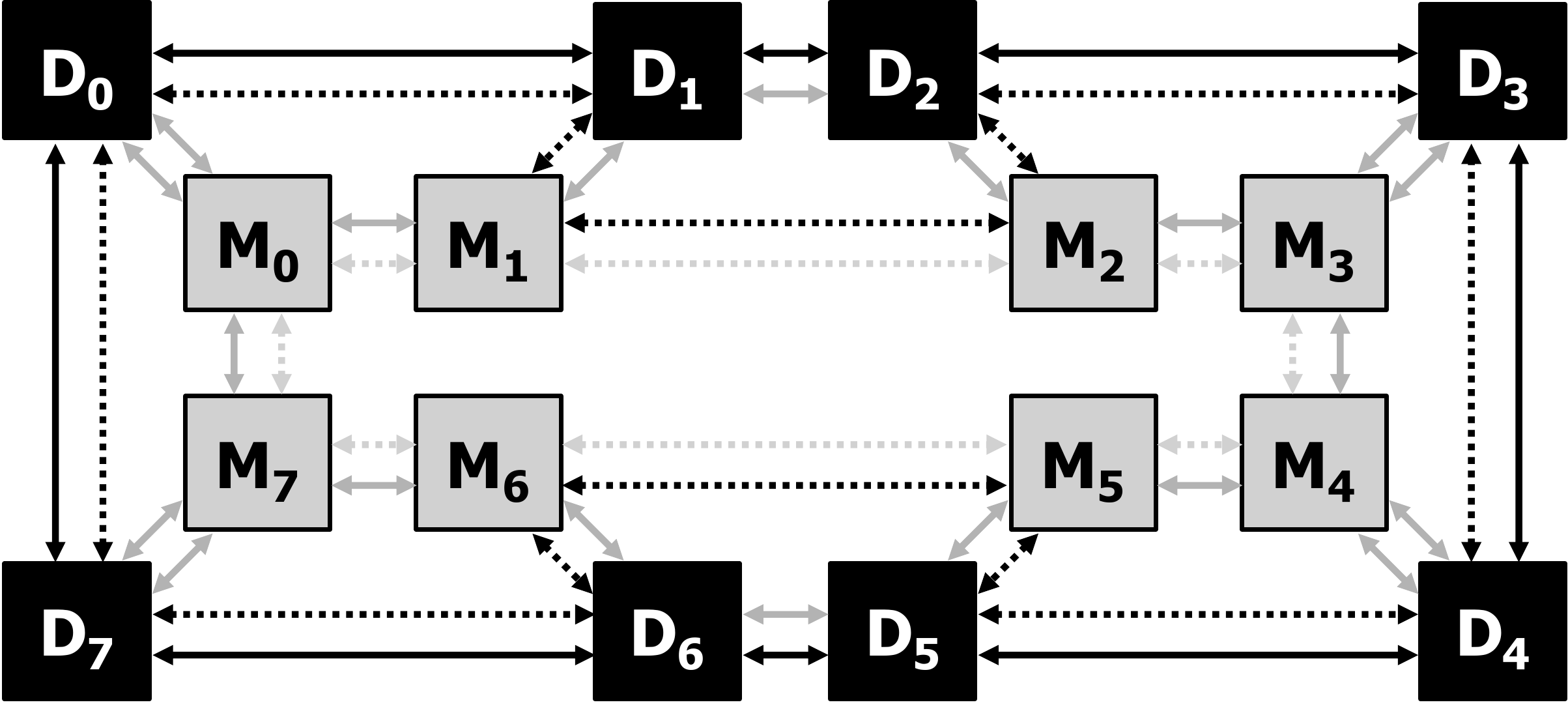}
	\label{fig:mcdla_m_star}
}
\vspace{1em}
\subfloat[]{
	\includegraphics[width=0.45\textwidth]{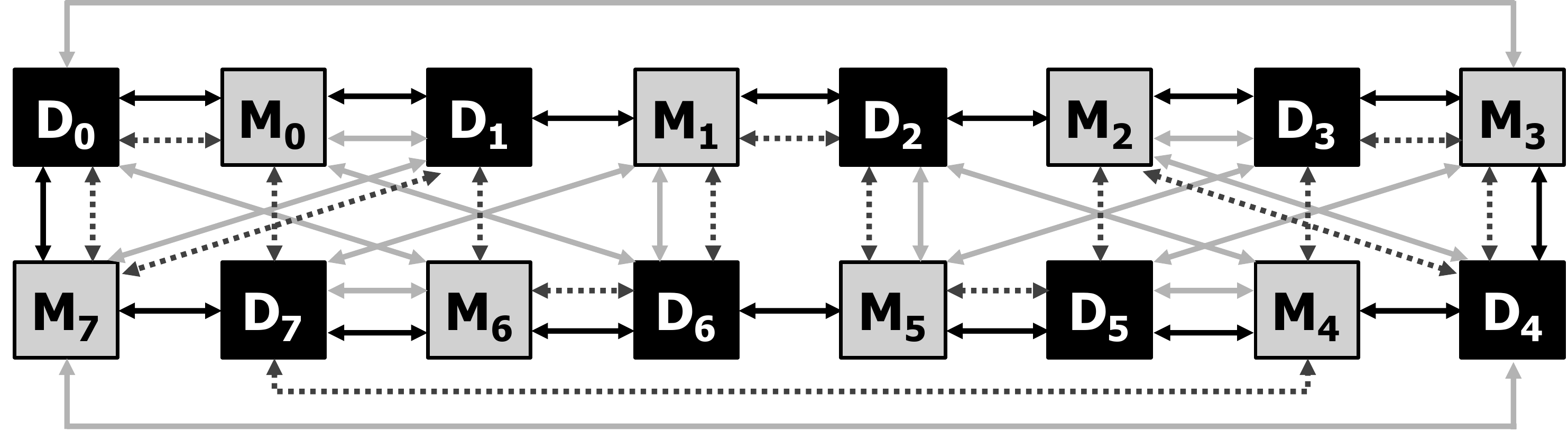}
	\label{fig:mcdla_ring}
}
\caption{
\mcdls system interconnect containing $8$ device-nodes and $8$ memory-nodes:
(a) a derivative interconnect design based on the cube-mesh topology of 
\fig{fig:dgx_1v}, where the device-nodes (\dnode{0-7}) communicate with
the memory-nodes (\mnode{0-7}) under a star topology, (b) the memory-nodes folded
inward, and (c) the proposed ring-based system interconnect.
Nodes that are part of the same ring are interconnected using the same color-coded
arrows (links).
}
\vspace{-1.2em}
\label{fig:mcdla_design_space}
\end{figure}

{\bf System Interconnect.} 
The design objective of the \mcdls device-side
interconnect is to balance communication, memory virtualization, and overall
design complexity.  A straightforward and an intuitive interconnect design that
can utilize our memory-nodes as a backing store to the device-nodes is shown in
\fig{fig:mcdla_design_space}(a).  Here, the communication links that constitute
one of the (\texttt{N}/$2$)=$3$ ring networks in \fig{fig:dgx_1v} (e.g., the
		singular ring network constructed using the $8$ bi-directional black
		arrows) are rearranged to construct a ring network using the $8$
memory-nodes and $8$ device-nodes. Each device-node is now provided with the
ability to access its designated memory-node using two high-bandwidth links
($50$ GB/sec communication bandwidth between
 \dnode{n}$\leftrightarrow$\mnode{n}), significantly reducing the latency to
migrate data to/from the backing store. There are two significant limitations 
with this design however, as (1) the $3$ rings used for inter-device collective
communication are constructed in a highly unbalanced fashion (i.e., $2$ rings
		are constructed with a maximum $8$ hop count while the remaining ring
		incurs a maximum $24$ hop count\footnote{In
		\fig{fig:mcdla_design_space}(a), each  memory-node is visited \emph{twice}
		when traversing the black-arrowed ring network, e.g.,
		$\cdots$\mnode{0}$\rightarrow$\dnode{0}$\rightarrow$\mnode{0}$\rightarrow$\mnode{7}$\rightarrow$\dnode{7}$\rightarrow$\mnode{7}
		$\cdots$}), rendering the overall communication latency be bottlenecked by
the longest ring, and (2) the $8$ light-gray/dotted bi-directional links are
neither being utilized for communication nor for memory virtualization, failing
to maximally utilize available communication resources\footnote{The
	light-gray/dotted arrows form the $4$\textsuperscript{th} ring with only the
		$8$ memory-nodes, without any device-nodes (i.e., all device-nodes are
				already fully utilizing the \texttt{N}=$6$ links). For parallel 
		DL training, the messages to be communicated across the devices are generated
		by the device-nodes (stored inside GPU memory) and never inside the
		memory-nodes, so the $4$\textsuperscript{th} ring does not help improve
		the performance of communication nor memory virtualization.}.
		\fig{fig:mcdla_design_space}(b) is an alternative design point that better
		balances the $3$ ring networks' performance (the $3$ rings containing the 
				device-nodes are constructed with a maximum $8$, $12$, and $20$ hop
				count, respectively), but it similarly suffers from the aforementioned
		limitations: unbalanced ring design and underutilization of communication
		resources.

\begin{figure}[t!] \centering
\subfloat[]{
\includegraphics[width=0.47\textwidth]{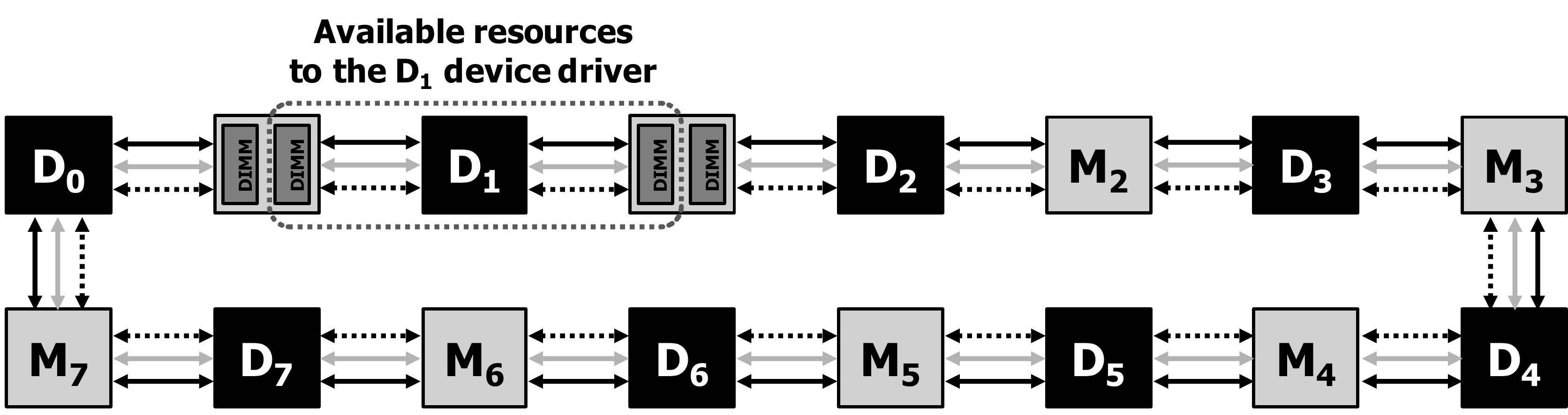}
}
\vspace{1em}
\subfloat[]{
\includegraphics[width=0.30\textwidth]{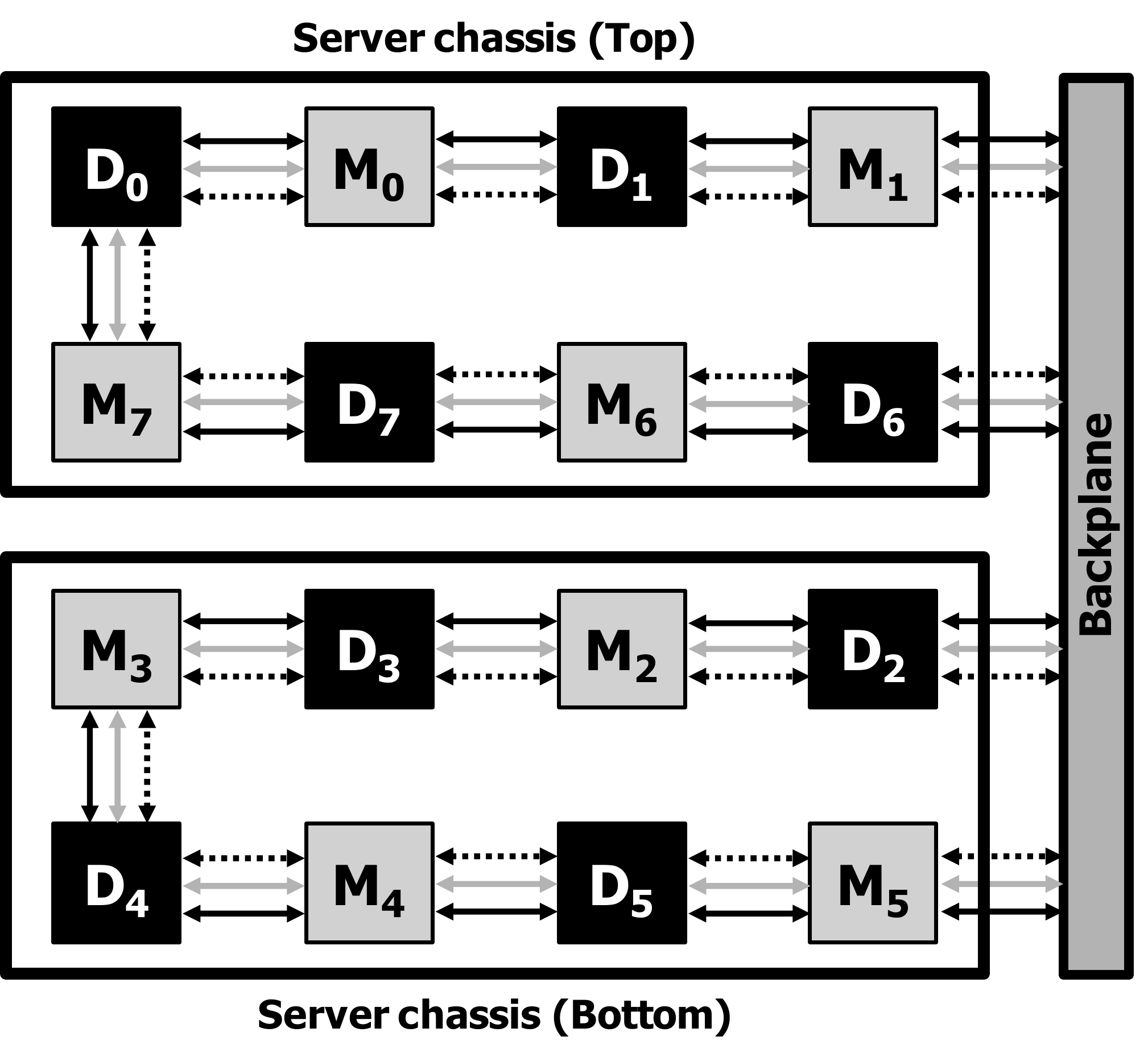}
}
\caption{
(a) Ring-based \mcdls system interconnect optimized for packaging 
 and design complexity (e.g., equal length inter-node links), and
 (b) the physical design of \mcdls where each half of the ring 
 is connected via the enclosure backplane.
}
\vspace{-1.2em}
\label{fig:mcdls_packaging_optimized}
\end{figure}

\begin{figure}[t!] \centering
\includegraphics[width=0.49\textwidth]{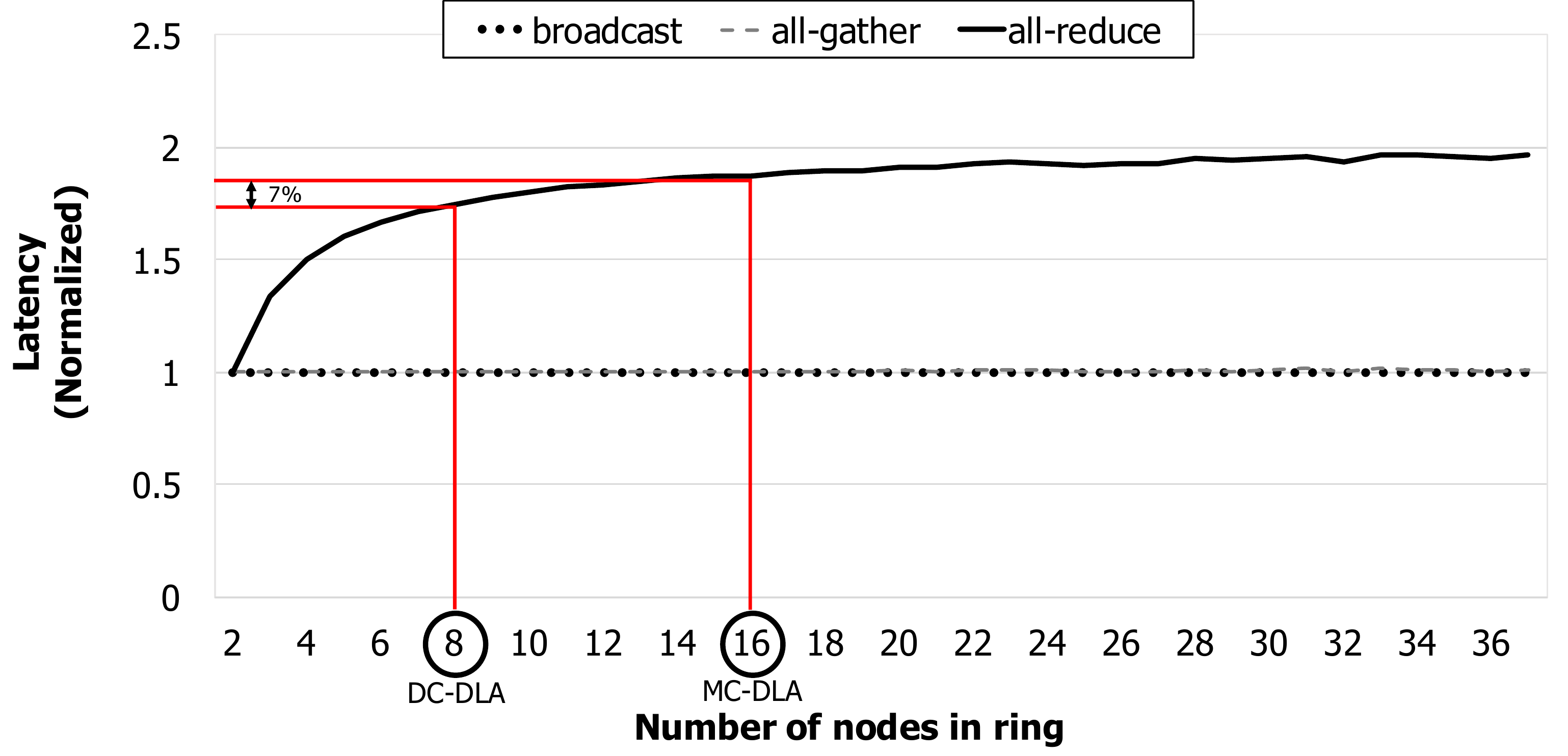}
\caption{
Latency incurred when performing collective communication primitives, as a function of the number of nodes inside
	the ring network (normalized to a ring with $2$ nodes). Each link has $50$ GB/sec of bi-directional bandwidth and
	the nodes communicate with a message size of $4$ KB with a target synchronization size at $8$ MB.
}
\label{fig:perf_collective}
\end{figure}

To holistically address these challenges, we arrive at our \emph{ring} based
\mcdls interconnect that maintains competitive collective communication
performance of \dcdls while, at the same time,
						significantly improving and maximizing the communication bandwidth
						available for memory virtualization.  \fig{fig:mcdla_design_space}(c)
						illustrates our proposed ring-based \mcdls interconnect design (\texttt{N/2}=$3$ rings overall),
						where  
	a given device-node is provided with a pair of high-bandwidth links to two
	memory nodes located on its left and right side of any given ring.
	The key advantage of our ring-based \mcdls architecture is twofold. First,
	each device-node is now able to utilize the two memory-nodes located on its
	(logical) left and right side of a given ring, maximally utilizing the
	\texttt{N}=$6$ links for virtualizing memory ($3\times$ higher bandwidth than
			in \fig{fig:mcdla_design_space}(a,b)).  This allows \mcdls to achieve (number
				of rings)*(link bandwidth to left and right nodes) =
			(\texttt{N}/$2$)*($2$*\texttt{B}) = $150$ GB/sec of communication
			bandwidth, a significant improvement over the legacy PCIe. Second, the
			communication bandwidth to the memory-nodes can linearly \emph{scale},
			proportional to the signaling technology used to implement these
			high-bandwidth links, as opposed to the PCIe-based \dcdls or \hcdls
			design, whose maximal communication bandwidth is capped at the maximum
			CPU socket-level memory bandwidth. For instance, both \dcdls and \hcdls,
			regardless of whether the host-device interface is designed using NVLINK
			or the next-generation PCIe, can only provide up to the maximum per CPU
			socket-level memory bandwidth, which is approximately $80$
			GB/sec and $120$ GB/sec for high-end Intel Xeon and IBM Power CPUs,
			respectively. The eight accelerator devices in our \mcdls has ($150$
					GB/sec per device $\times$ $8$ devices) = $1200$ GB/sec of
			communication bandwidth to its neighboring memory-nodes, the number of
			which will proportionally grow as a function of the link bandwidth of
			\texttt{B} GB/sec. \fig{fig:mcdls_packaging_optimized} is an illustration 
			of our ring-based \mcdls re-designed to be optimized for packaging costs,
			easing its adoption for real-world HPC systems.

One might be concerned that the latency incurred for collective communications
will increase as \mcdls adds $8$ additional (memory) nodes,
					 effectively doubling the number of nodes inside the ring.  
For reasonably large messages, our ring-based \mcdls with
					 $16$ nodes incur negligible latency overheads for all-gather,
					 broadcast, and all-reduce (\fig{fig:perf_collective}). 
					 When the communication size is small, \mcdls does incur higher latency 
					than \dcdls, but in such scenario the communication latency is not a 
performance limiter to begin with (Amdahl's law).					 
					 We demonstrate in
					 \sect{sect:results} that the impact of this latency overhead is
					 negligible on system performance.

{\bf Software Interface.} 
\mcdls builds upon the memory-overlaying
based DNN memory virtualization solutions~\cite{rhu:2016:vdnn,ibm_lms}, which assume the following: (1) the high-level DL framework
analyzes the neural network DAG structure at compile-time and derives the
data-dependencies of memory-hungry DNN data, (2) this information is utilized
by the runtime memory manager to schedule performance-aware, software-managed 
memory overlaying operations (i.e., DMA-initiated \texttt{cudaMemcpyAsync}) across 
host-device memory to expand the reach of memory available
for training.  The \mcdls design introduces another tier of memory region 
in addition to the host and device memory -- the capacity-optimized
memory inside the memory-nodes, which we refer to as \deviceremote memory	in
the rest of this paper. We propose to utilize \deviceremote memory to supplant
the role of host memory for stashing DNN data with long reuse distance. 
In other words, memory virtualization is implemented using the local
device (\devicelocal) memory and \deviceremote memory without having the CPU memory involved.
To allow the runtime memory manager to (de)allocate data structures inside 
\deviceremote memory and initiate DMA data transfers in/out of this memory region, 
we introduce three extensions to the CUDA runtime APIs (\texttt{libcudart.so})
for \deviceremote memory (de)allocation and 
memory copy (\tab{tab:api}).
Using these APIs, existing DL frameworks can seamlessly exploit
the additional pool of memory inside our memory-nodes.

\begin{table}[t!]
  \centering
  \caption{Software API extensions for \mcdls. }
%\scriptsize
\footnotesize
\vspace{-0.5em}
\resizebox{0.485\textwidth}{!}{%
  \begin{tabular}{c c c}
    \hline
    \texttt{API} & \texttt{Arguments} & \texttt{Semantics} \\
    \hline
    \hline
   	\texttt{cudaMallocRemote}	& \texttt{\&src, size}	& malloc \texttt{size} bytes to \deviceremote\\
															&												& memory and return ptr to \texttt{src}\\
    \hline                              
    \texttt{cudaFreeRemote}		& \texttt{\&src}				& free memory that is allocated under \\
															&												& \deviceremote memory \\
		\hline
   	\texttt{cudaMemcpyAsync}	& \texttt{\&src, \&dst,}	& copy \texttt{size} bytes from \texttt{src} to \texttt{dst},\\
															&	\texttt{size,}					& but \texttt{direction} now includes \\
															& \texttt{direction}			& \texttt{LocalToRemote} and \texttt{RemoteToLocal}\\
		\hline
  \end{tabular}%
  }
\vspace{-1.2em}
  \label{tab:api}
\end{table}

\begin{figure}[t!] \centering
\includegraphics[width=0.37\textwidth]{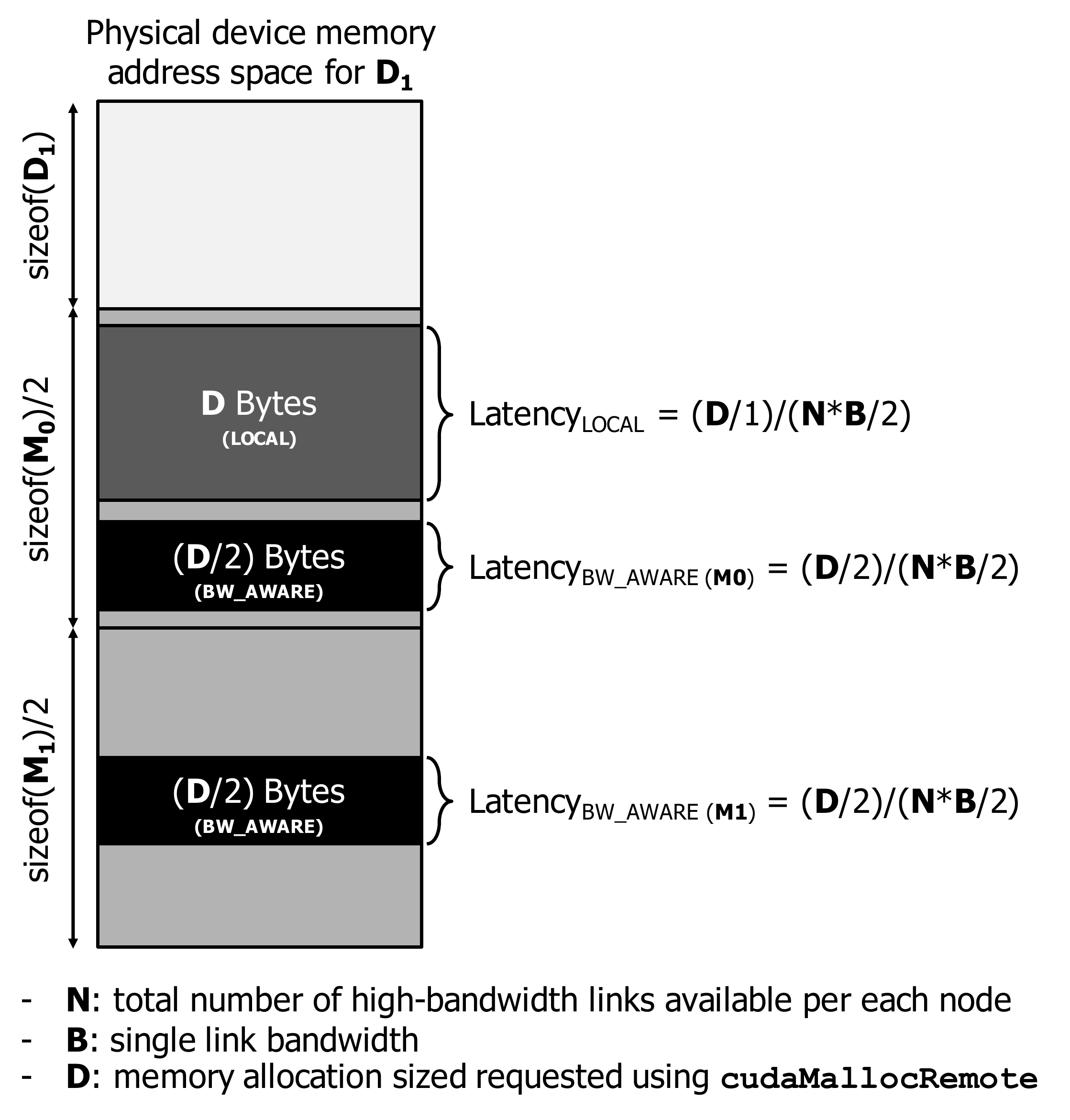}
\caption{
The \texttt{LOCAL} and \texttt{BW\_AWARE} page allocation policy employed in \mcdls. \texttt{BW\_AWARE}
allows the device-node \texttt{D$_{1}$} to read (write) data from (to) the left and right memory-nodes
	concurrently, reducing the overall latency by half compared to \texttt{LOCAL}.
}
\vspace{-1.2em}
\label{fig:bw_aware}
\end{figure}

{\bf System Software Support.} \mcdls requires the device driver to be
able to (de)allocate memory in \deviceremote memory and be able to map that
address space to user-level programs. Under our design, 
				any given memory-node is logically partitioned into two 
	groups and all the resources within a group (e.g., DMA engine, memory
			controller, and memory DIMMs) are exclusively assigned to a single
	device-node for servicing its requests (\fig{fig:mcdls_packaging_optimized}). As these resources are not
	to be shared by any two device-nodes by design, the device driver manages
	both its client device-node and the each half of the left and right side
	memory-nodes's physical memory under a single device memory address space.
	Consequently, the \devicelocal physical memory lives at the bottom of this
	single device memory address space and each half of the two \deviceremote
	physical memory is concatenated and mapped into the higher address space
	(\fig{fig:bw_aware}). From the driver's perspective, the device-node
	augmented with its share of memory-nodes can be thought of as a single PCIe device
	but with a larger memory capacity (e.g., Maxwell M40 containing $12$ GB
			versus Volta V100 with $16$ GB), hence existing system software APIs
	(e.g., \texttt{mmap}) can be used as-is to map the enlarged device memory
	address region to the user-level space. The current design of \mcdls can  
	add up to $1.3$ TB $\times$ $8$ = $10.4$ TB of
	additional physical memory (\sect{sect:power}), well fitting under the
	addressing capabilities of current GPUs (e.g., $49$-bit virtual addressing
			($512$ TB) and $47$-bit physical memory addressing ($128$ TB))~\cite{pascal_p100}. The added memory capacity to each
	device-node is informed to the device driver at boot-time so that the driver
	takes it into consideration when (de)allocating memory. Allocating pages in
	both \devicelocal and \deviceremote memories can be done using existing
	device-side page-tables and the page-table walker, but our page
	allocation/placement policy is designed in a \emph{bandwidth-aware} (\texttt{BW\_AWARE}) manner in
	order to maximally exploit the high-bandwidth communication channels to the
	left and right memory nodes. Consider a
\texttt{cudaMallocRemote} call with $D$ Bytes of memory allocation requested to
the driver. Rather than having the entire $D$ Bytes of data be allocated under a single
memory-node (which we refer to as \texttt{LOCAL} allocation policy\footnote{
The \texttt{LOCAL} allocation policy is named after the \emph{local} NUMA zone page allocation policy of
\texttt{libNUMA} in Linux and is not intended to imply that allocation is done inside \devicelocal memory.
}), 
	our proposal splits the requested malloc size into two equal sized chunks (aligned in page granularity) 
	and maps the pages within each chunk to the left and right memory-node's share of the memory address space in 
a round-robin fashion. This allows the device-node to utilize all \texttt{N}
high-bandwidth links to read/write data from the two memory-nodes, maximally utilizing the
\texttt{N}$\times$\texttt{B} GB/sec of memory bandwidth for memory virtualization.

\begin{figure*}[t!] 
\centering
\subfloat[]{
	\includegraphics[width=0.80\textwidth]{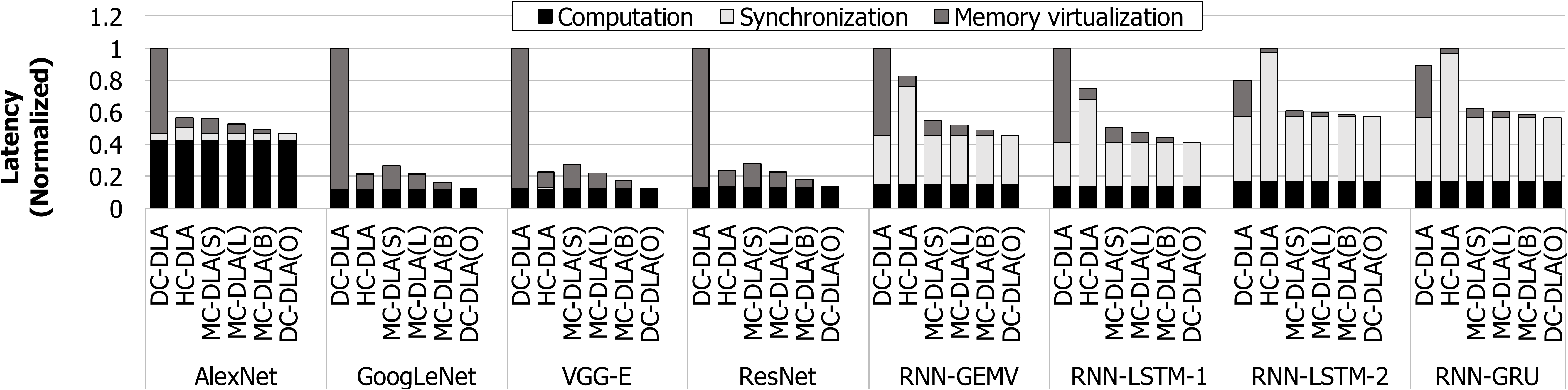}
	\label{fig:latency_breakdown_data_parallel}
}
\vspace{1em}
\subfloat[]{
	\includegraphics[width=0.80\textwidth]{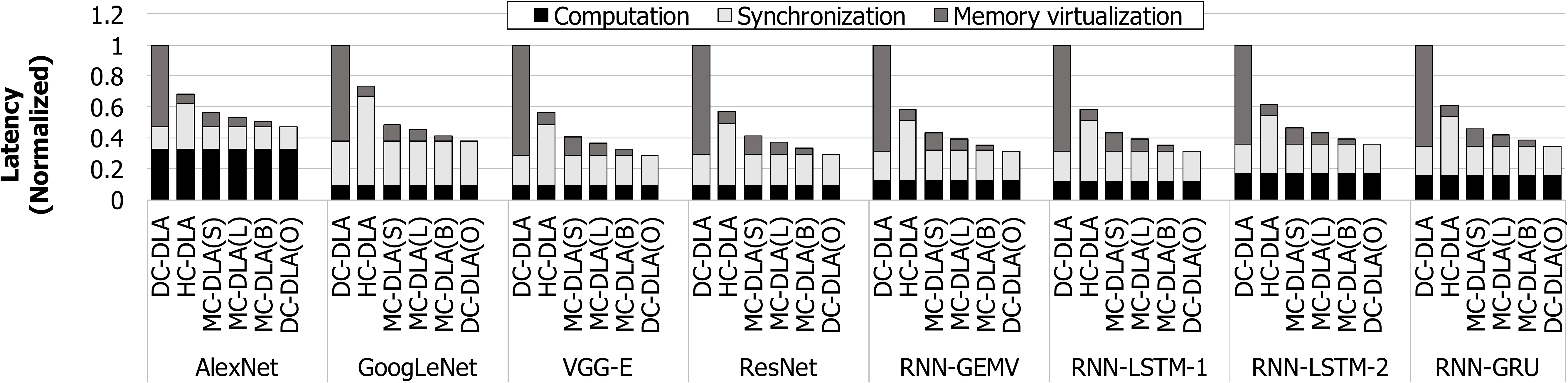}
	\label{fig:latency_breakdown_model_parallel}
}
\caption{
Breakdown of latencies incurred during: (a) data-parallel training, (b) model-parallel training. Figures are normalized to the highest
	stacked bar chart. Note that the sum of these three latency categories does not directly translate into the system-level performance because DL frameworks try to overlap computation time with synchronization and memory virtualization.
}
\vspace{-1.2em}
\label{fig:latency_breakdown}
\end{figure*}

\begin{table}[t!]
  \centering
  \caption{Device-/memory-node configuration parameter. }
%\small
\footnotesize
%\scriptsize
  \begin{tabular}{|c|c|}
		\hline
		\multicolumn{2}{|c|}{\textbf{Device-node}} \\
		\hline
    Number of PEs     			& $1024$   \\
    \hline               
    MACs per PE   					& $125$ \\
    \hline              
    PE operating frequency				& $1$ GHz \\
    \hline              
    Local SRAM buffer size per PE   			& $32$ KB \\
    \hline              
    Memory bandwidth   			& $900$ GB/sec \\
    \hline              
    Memory access latency   			& $100$ cycles  \\
		\hline
		Number of high-bandwidth links (\texttt{N})	& $6$ \\
		\hline
		Communication bandwidth per link (\texttt{B})	& $25$ GB/sec \\
    \hline             
 
		\multicolumn{2}{|c|}{\textbf{Memory-node}} \\
		\hline
		Memory bandwidth	& $256$ GB/sec	\\
		\hline
		Memory access latency & $100$ cycles	\\
		\hline
		Number of high-bandwidth links (\texttt{N})	& $6$ \\
		\hline
		Communication bandwidth per link (\texttt{B})	& $25$ GB/sec \\
		\hline
  \end{tabular}
\vspace{-1.2em}
  \label{tab:d_node_config}
\end{table}

\section{Methodology}
\label{sect:eval}

{\bf Device and Memory Nodes.} We developed an in-house system-level simulator
for evaluating \mcdls. 
The high-level architecture design of our DL accelerator resembles that of
Eyeriss~\cite{eyeriss_isca} or DaDianNao~\cite{dadiannao} in that our device
architecture also employs a spatial array of processing elements (PEs), each of
which contains (1) a multitude of MAC operators for handling vector operations,
			(2) local SRAM buffers (double-buffered to overlap computation with data
					fetches) to leverage data locality, and (3) a high-bandwidth
			on-package memory (e.g., HBM~\cite{hbm}) for local \devicelocal allocations.
			The baseline device-node has been configured as summarized in \tab{tab:d_node_config} but 
			we also evaluate \mcdls's sensitivity to alternative configurations in \sect{sect:perf}.  Our model is
			designed to optimize generic GEMM (general matrix multiplication)
	operations so that it handles not only convolutional layers, but also
	recurrent layers, fully-connected layers, activation layers, and etc. Based
	on our analysis, an output-stationary dataflow (i.e., output feature maps are
			stationed locally on-chip) as discussed by Chen et
	al.~\cite{eyeriss_isca} provides a good balance in terms of MAC utilization
	and energy-efficiency across all of the layers we evaluate, hence our device
	accelerator employs the output-stationary dataflow rather than the
	row-stationary dataflow. It is worth pointing out that the scope of this
	paper is on studying HPC system architectures for DL training, rather than
	the development of a high-efficiency accelerator device. Therefore, our proposal
	is equally effective for alternative DL accelerator designs and
	DNN dataflows.  Note that a single iteration of training can take hundreds of
	milliseconds even on a real high-end GPU card, so being able to perform
	simulation in tractable amount of time is crucial. We therefore model the
	device-node's HBM memory and the memory DIMMs inside the memory-nodes as
	having fixed memory bandwidth and latency, rather than resorting to a
	cycle-level DRAM simulator~\cite{dramsim2,usimm,ramulator}. We believe our
	methodology provides accurate estimations without losing fidelity due to the
	following two reasons: (1) as DNN computation and memory accesses have high
	data locality with highly deterministic dataflow, existing
	designs~\cite{dadiannao,eyeriss,scnn} primarily employ a lightweight FSM or
	microcontrollers to orchestrate on and off-chip data movements in
	coarse-granular data sizes and (2) all inter-node (e.g.,
			\texttt{host}--\devicelocal, \devicelocal--\deviceremote) data copy
	operations are conducted as coarse-grained, bulk data transfers using DMAs
	(\sect{sect:vdnn}) with high data locality, rendering the system-level
	performance being less sensitive to the underlying behavior of the DRAM
	microarchitecture (e.g., bank conflicts).

\begin{table}[t!]
  \centering
  \caption{Evaluated benchmarks. }
%\small
\footnotesize
%\scriptsize
  \begin{tabular}{|c|c|c|}
    \hline
    \textbf{Network} & \textbf{Application} &  \textbf{\# of layers} \\
    \hline
    \hline
    AlexNet     			& Image recognition &  $8$	\\
    \hline               
    GoogLeNet   			& Image recognition &   $58$\\
    \hline               
    VGG-E   					& Image recognition &   $19$	\\
    \hline               
    ResNet   					& Image recognition &   $34$	\\
    \hline              
		\hline 
    \textbf{Network} & \textbf{Application} &  \textbf{Timesteps} \\
    \hline
    \hline
    RNN-GEMV	& Speech recognition  &  	$50$ \\
    \hline             
    RNN-LSTM-1	& Machine translation & $25$ \\
    \hline             
    RNN-LSTM-2	& Language modeling  &   $25$ \\
    \hline               
    RNN-GRU		& Speech recognition  &  $187$ \\
    \hline
  \end{tabular}
\vspace{-1.2em}
  \label{tab:benchmarks}
\end{table}

{\bf System Architecture.} We assume an $8$ device-node system configuration in
all of our experiments (\fig{fig:dcdls_vs_mcdls}). The baseline \dcdls system
architecture is modeled after NVIDIA's DGX system~\cite{dgx_1v} and IBM's
PowerAI DDL system~\cite{ibm_ddl}. Both of these HPC systems employ the
cube-mesh device-side interconnect, which is flattened into multiple ring
networks (three rings in our evaluation, \texttt{N}=$6$ links per device node,
		\sect{sect:sys_arch}) to maximally utilize inter-node link bandwidth and
minimize the latency incurred in conducting inter-device
communications~\cite{chan2006collective,nccl:wooley}. \dcdls uses PCIe (gen3)
to communicate with the host memory for memory virtualization.  The \hcdls
system architecture is modeled after IBM-NVIDIA's Power9
Summit~\cite{ibm_summit}, which assumes the following: (1) among the \texttt{N}
high-bandwidth links available to each device-node, \hcdls allocates half of
them to be connected to the CPU memory for reads and writes, trading off fast
memory virtualization over communication in a \emph{balanced} manner, (2) four
device-nodes are connected to a single CPU socket (i.e., $8$ devices sharing two
		sockets), and (3) the maximum CPU socket memory bandwidth is large enough
to fully service the aggregate CPU memory bandwidth usage of the four
device-nodes that are connected to that CPU socket.  Consequently, this
hypothetical CPU in \hcdls has $300$ GB/sec of per socket CPU memory bandwidth ($3\times$ to $4\times$ overprovisioned than
		real systems~\cite{intel_xeon,ibm_power9})
						 which allows half of the $N$(=6) high-bandwidth links to be used
						 to read/write CPU memory (i.e., $4$$\times$\texttt{B}$\times3$ =
								 $300$ GB/sec).  As \hcdls can consume up to $100\%$ of the
						 provisioned CPU memory bandwidth (we discuss the maximum and
								 average CPU memory bandwidth usage of all our system design
								 points in \sect{sect:bottlenecks}), such high CPU memory
						 bandwidth usage could potentially incur destructive interference
						 on CPU's role~\cite{purestorage_airi} in the overall DL training
						 process (e.g., running the DL framework software, interacting with
								 the backing storage HDD/SSD to fetch training data, etc),
						 slowing down the overall training time.  For a conservative
						 evaluation, we assume that \hcdls's CPU memory bandwidth usage has
						 \emph{no} effect on system performance.  We omit the results of
						 \hcdls designs that partitions high-bandwidth links in an
						 asymmetric manner as these design points were shown to be less
						 robust than the studied, balanced \hcdls.  An \emph{oracular}
						 version of \dcdls was also established by having an infinitely
						 sized on-package, \devicelocal memory available inside each device-node,
						 obviating the need for CPU-GPU data migration.  We explore such
						 (unbuildable) system design point to evaluate the effectiveness of
						 \mcdls.  The memory-nodes in \mcdls are configured to house ten
						 DDR4 DIMMs providing a maximum of $256$ GB/sec of memory bandwidth
						 to the neighboring device-nodes.

{\bf Benchmarks.} 
We study a diverse set of eight DNN applications (\tab{tab:benchmarks}) that
encompasses not only convolutional neural networks (CNNs) but also recurrent
neural networks (RNNs). 
We choose four CNN topologies that show
state-of-the-art performance in ImageNet~\cite{imagenet}, namely AlexNet,
	GoogLeNet, VGG-E, and ResNet. The four RNN applications have been chosen from
	Baidu's DeepBench application suite~\cite{deepbench}, which includes one GEMV-based vanilla
	RNN topology, two LSTM-based, and one GRU-based RNNs. We use a batch size of
	$512$ for all our evaluations and study both data-parallel and model-parallel
	training (\fig{fig:parallel_dl_training}) for partitioning the DL algorithm
	across the eight device-nodes. For model-parallel training, we employ the model-parallelization 
	strategy as employed by Krizhevsky et al.~\cite{alexnet}.

{\bf Memory-overlaying for DNN Virtual Memory.} We implemented the runtime
memory management policy as described in
\cite{rhu:2016:vdnn,ibm_ddl,ibm_lms,lms_sysml}, which leverages the network DAG
to analyze inter-layer data dependency to schedule memory-overlaying operations
for virtual memory. Under our implementation, the device memory is utilized as
an application-level cache with respect to the host memory. Concretely, the
runtime memory manager pushes all layer's feature maps to the backing store
after its last reuse during forward propagation and prefetches them back to the
local device memory during backpropagation\footnote{ We employ one exception to
	this rule: for layers that have short computation time (e.g., activation
			layers, pooling layers, $\ldots$), the memory manager chooses to
		\emph{re-compute} the feature maps during backpropagation rather than
		migrating these data to the backing store. Such optimization minimizes the
		number of memory overlaying operations and is currently employed in
		MXNet~\cite{mxnet,chen:2016:sublinear_cost}. We adopt such optimization for
		a conservative evaluation and make sure the system performance is not
		unnecessarily degraded.  }. While some of these
		local$\leftrightarrow$remote data migration operations might not be
		necessary for DNNs that fit within the memory capacity limits, following
		prior work~\cite{rhu:2016:vdnn,park:2018:ppopp,lms_sysml,nikolai:gtc:2017},
		we employ such memory management policy to maximally stress the system
		interconnect. In other words, the $8$ DNN applications we study are used as
		microbenchmarks to stress test the system interconnect and evaluate its
		robustness in providing a performant virtual memory system without compromising
		communication performance.

%%%%%%%%%%%%%%%%%%%%%%%%%%%%%%%%%%%%%%%%%%%%%%%%%%%%%%%%%%%%

\section{Evaluation} 
\label{sect:results}

This section evaluates six system design points, the baseline \dcdls, a hypothetical
\hcdls design (\sect{sect:eval}), one star-topology based \mcdls
(\fig{fig:mcdla_design_space}(b), \mcdlsS), two ring-based \mcdls design points
(with \texttt{LOCAL} and \texttt{BW\_AWARE} page allocation policy, denoted as
 \mcdlsL and \mcdlsB, respectively), and an oracular \dcdls with infinite
memory size (\texttt{DC-DLA(O)}). All average values are based on harmonic means.
	
	 \begin{figure}[t!] \centering
\includegraphics[width=0.49\textwidth]{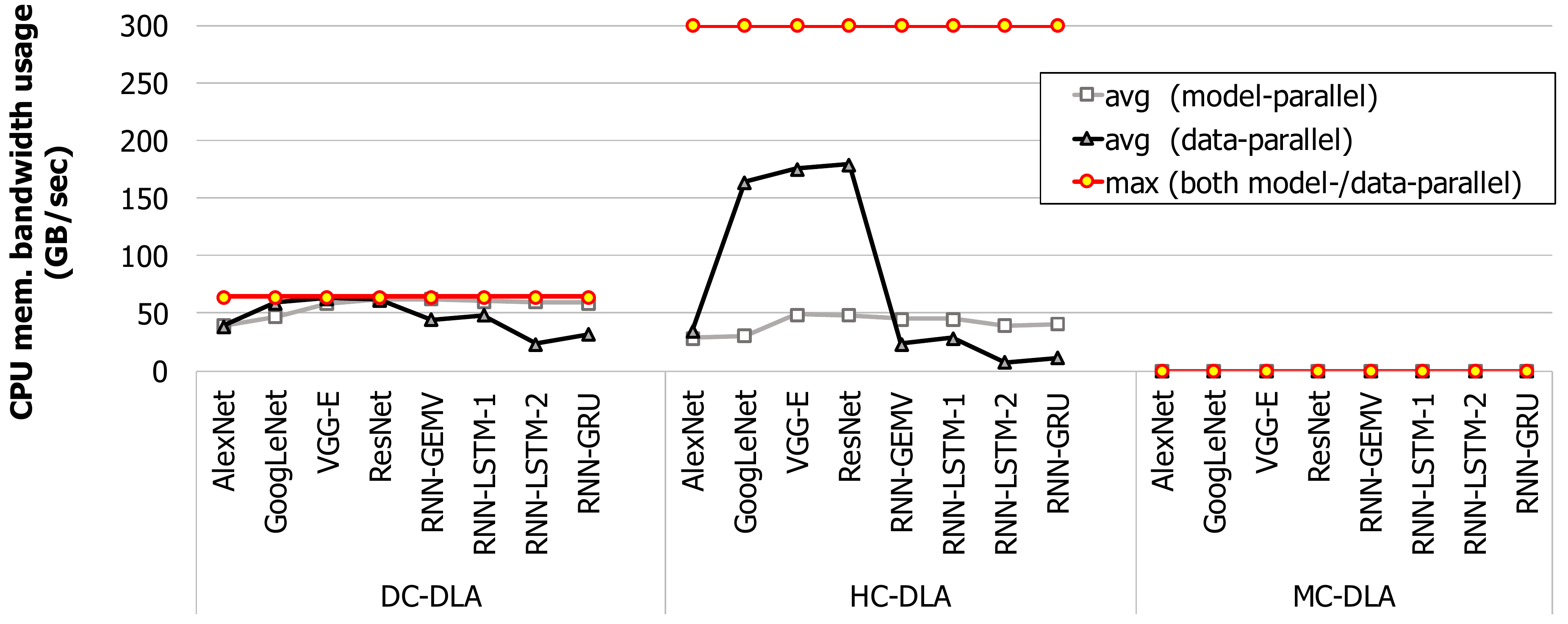}
\caption{
CPU memory bandwidth usage under different \dls designs.
}
\vspace{-1.2em}
\label{fig:cpu_mem_bw_usage}
\end{figure}

\subsection{Identifying System Bottlenecks}
\label{sect:bottlenecks}

Convolutional layers are generally compute-limited (e.g., sliding window
		based dataflow manifests high data locality) and its feature
maps, rather than weights, dominate memory allocation during 
training. Conversely, fully-connected layers and recurrent layers are 
memory bandwidth-limited where its weights take up a larger fraction of
the memory allocation than feature maps. Consequently, data-parallel
training of CNNs are generally insensitive to the underlying system's ability to
provide fast inter-device communication because the synchronization data size
(i.e., size of the weight gradients, \gradientW) is relatively much smaller
than its feature map size. Memory virtualization can therefore become a
performance bottleneck for data-parallel training of CNNs. RNNs however have a
relatively larger \gradientW size for synchronization
hence both fast communication and high-bandwidth memory
virtualization is required for data-parallel RNN training. Model-parallel
training, as discussed in \sect{sect:parallel_dl_training}, incurs much
frequent (and larger) synchronization operations  than its data-parallel
counterparts, so a high-bandwidth device-side interconnect is crucial
for scalable DL training.

In this context,  to clearly illustrate the system-level performance
bottlenecks, we derive the latencies incurred in performing the (a)
	computations required for forward and backward propagation, (b) inter-device
	synchronization, and (c) memory-overlaying for memory virtualization, the
	three of which are stacked altogether in a single bar chart as shown in
	\fig{fig:latency_breakdown}. Overall, \dcdls spends the least amount of time
	on synchronization thanks to its high-bandwidth device-side interconnection
	network.  Memory virtualization however causes a significant performance
	bottleneck for \dcdls on $14$ out of the $16$ training examples because the
	PCIe links only provide a small fraction of what the high-bandwidth links can
	service.  Thanks to the high-bandwidth links allocated to access CPU memory,
	\hcdls can significantly reduce the latency incurred in memory virtualization
	(average $88\%$ reduction). This however comes at a cost where: (1) the
	smaller inter-device communication bandwidth now leads to an increase in
	synchronization time (average $90\%$ increase), and (2) the multiple devices
	in \hcdls that are utilizing CPU memory for virtual memory are now consuming
	a significant fraction of CPU memory bandwidth as shown in
	\fig{fig:cpu_mem_bw_usage}.  For DL training, CPUs play the role of (1)
	running the DL framework software, and (2) getting the training datasets
	ready to be fed into the accelerator devices (e.g., reading and batching
			input datasets from the HDD/SSD storage devices, preprocessing the input
			batches, and uploading the preprocessed input batches to the CPU
			memory)~\cite{purestorage_airi}.  While it is possible that such high
	host-device interaction can cause destructive interference and cause
	performance slowdown on \hcdls, for a conservative analysis, we do not take
	such behavior into account when evaluating overall performance in
	\sect{sect:perf}.  The three \mcdls designs are able to achieve the best of both \dcdls and \hcdls
		as the system interconnect successfully reduces the latency incurred in virtualizing memory while not
			having to incur noticeable overhead in inter-device communications. Additionally, because
		memory virtualization is provided using \deviceremote memory, there are no
		CPU memory bandwidth consumption whatsoever, enabling a system design that
		scales independently to its ties with the host interface. 

\begin{figure}[t!] 
\centering
\subfloat[]{
	\includegraphics[width=0.49\textwidth]{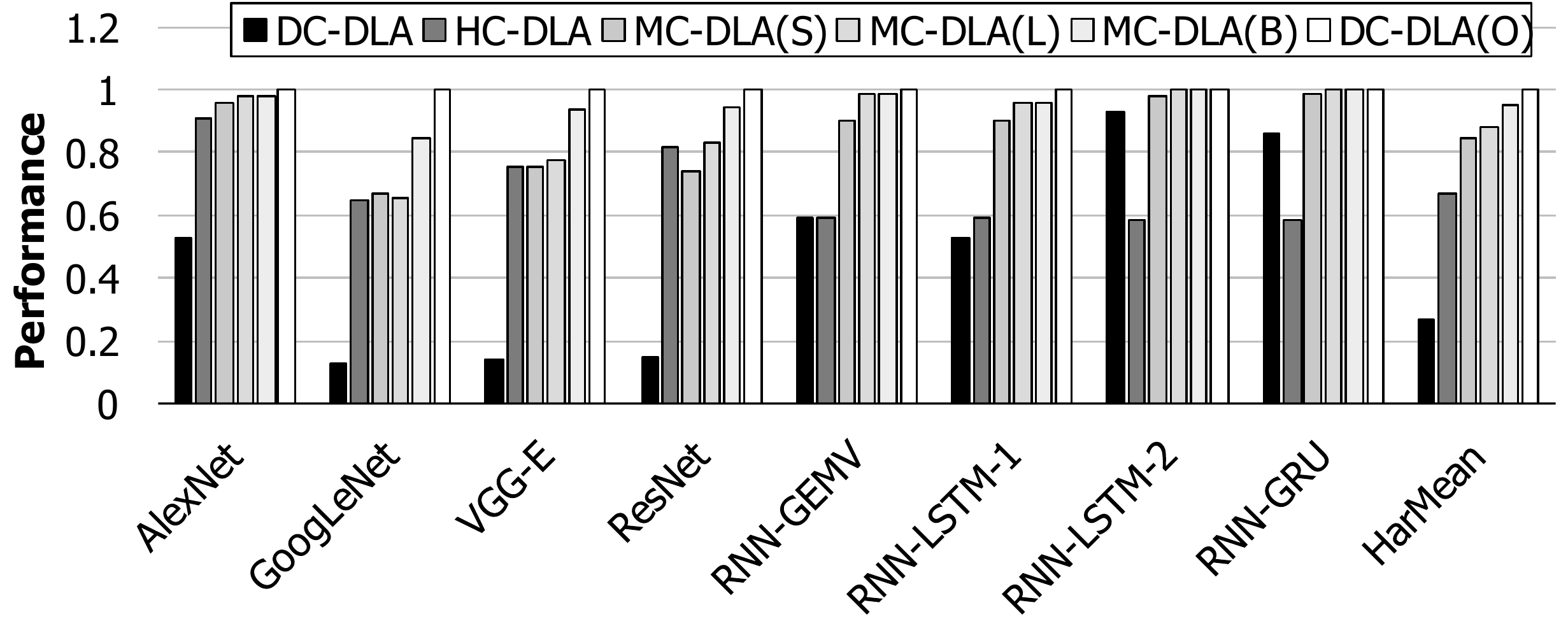}
	\label{fig:perf_data_parallel}
}
\vspace{1em}
\subfloat[]{
	\includegraphics[width=0.49\textwidth]{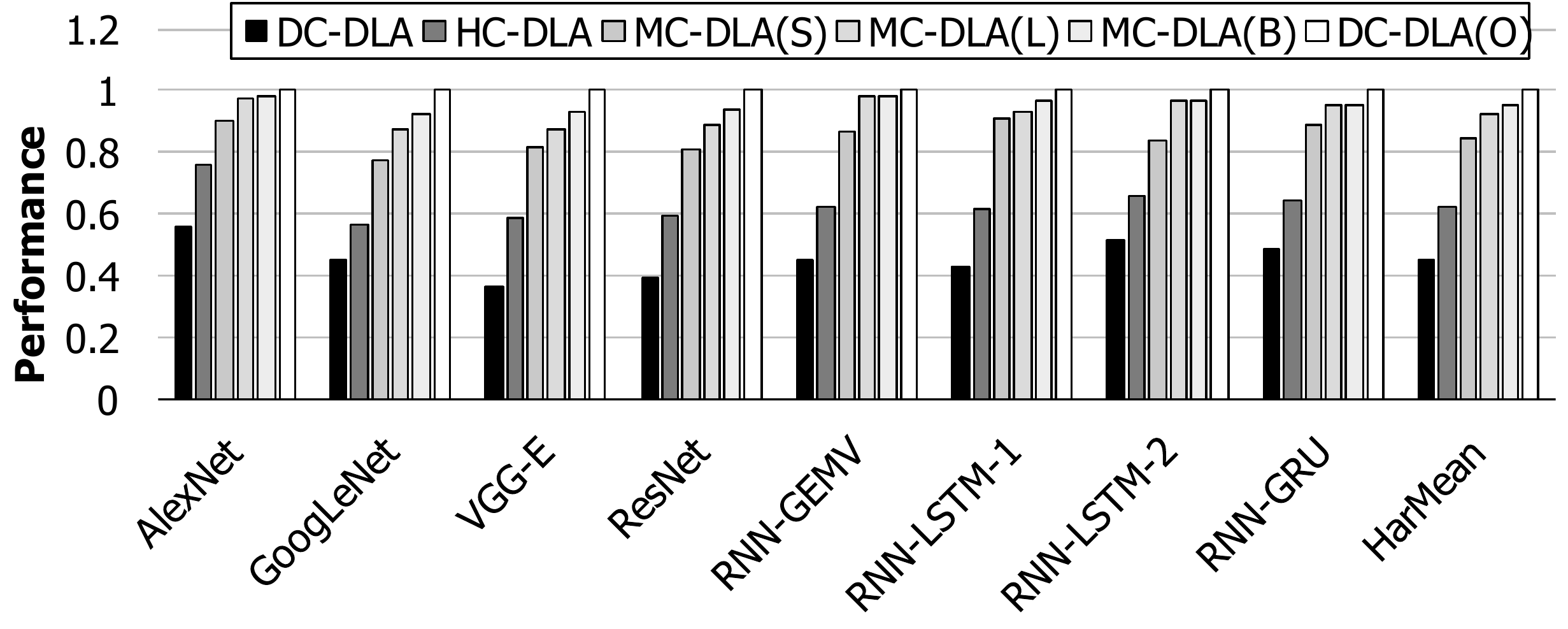}
	\label{fig:perf_model_parallel}
}
\caption{
Performance improvements offered by \mcdls for: (a) data-parallel training, (b) model-parallel training.
}
\label{fig:perf}
\end{figure}

%------------------------------
\subsection{Performance}
\label{sect:perf}

\fig{fig:perf} summarizes the performance of \mcdls compared to \dcdls,
\hcdls, and the oracular \dcdls.  The \hcdls design provides an average $32\%$ and $38\%$ speedup over
\dcdls for data-parallel and model-parallel training, respectively. This is due
to \hcdls's ability to balance fast communication and memory virtualization,
	 which \dcdls fails in achieving due to its asymmetric partitioning of
	 communication bandwidths (i.e., more than $10\times$ difference in bandwidth
			 provisioned for inter-device communication and memory virtualization).
	 \hcdls however is only able to leverage half of its high-bandwidth links for
	 communication and virtual memory, failing to maximally benefit from the device-side
	 interconnect.  Our proposed \mcdlsB design fully unlocks the \texttt{N}
	 high-bandwidth links  for both communication and memory virtualization,
	 leading to an average $3.5\times$ and $2.1\times$ speedup over \dcdls for
	 data-parallel and model-parallel training, respectively (average $2.8\times$). 
	 Moreover, \mcdlsB reaches $84\%$--$99\%$ of the performance of an unbuildable,
	 oracular \dcdls (average $95\%$). While \mcdlsS does much better than \dcdls or \hcdls,
	 its suboptimal utilization of high-bandwidth links leaves significant performance left on
	 the table (maximum $24\%$, average $14\%$ performance loss than \mcdlsB).
	 It is worth
	 pointing out that the relatively sub-optimal, but simpler \mcdlsL design
	 achieves $96\%$ of the performance of \mcdlsB. Although \mcdlsL is only
	 provisioned with half the memory virtualization bandwidth of \mcdlsB, the
	 high-bandwidth communication channels for synchronization are equally
	 provided for both designs thanks to its ring-based
	 system interconnect.  While \mcdlsL and \mcdlsB provides similar benefits
	 over the $8$ applications we study in this paper, we believe \mcdlsB to be a
	 more robust and scalable design option as it can maximally utilize the
	 interconnect bandwidth with reasonable design costs (\sect{sect:sys_arch}).

\begin{figure}[t!] \centering
\includegraphics[width=0.49\textwidth]{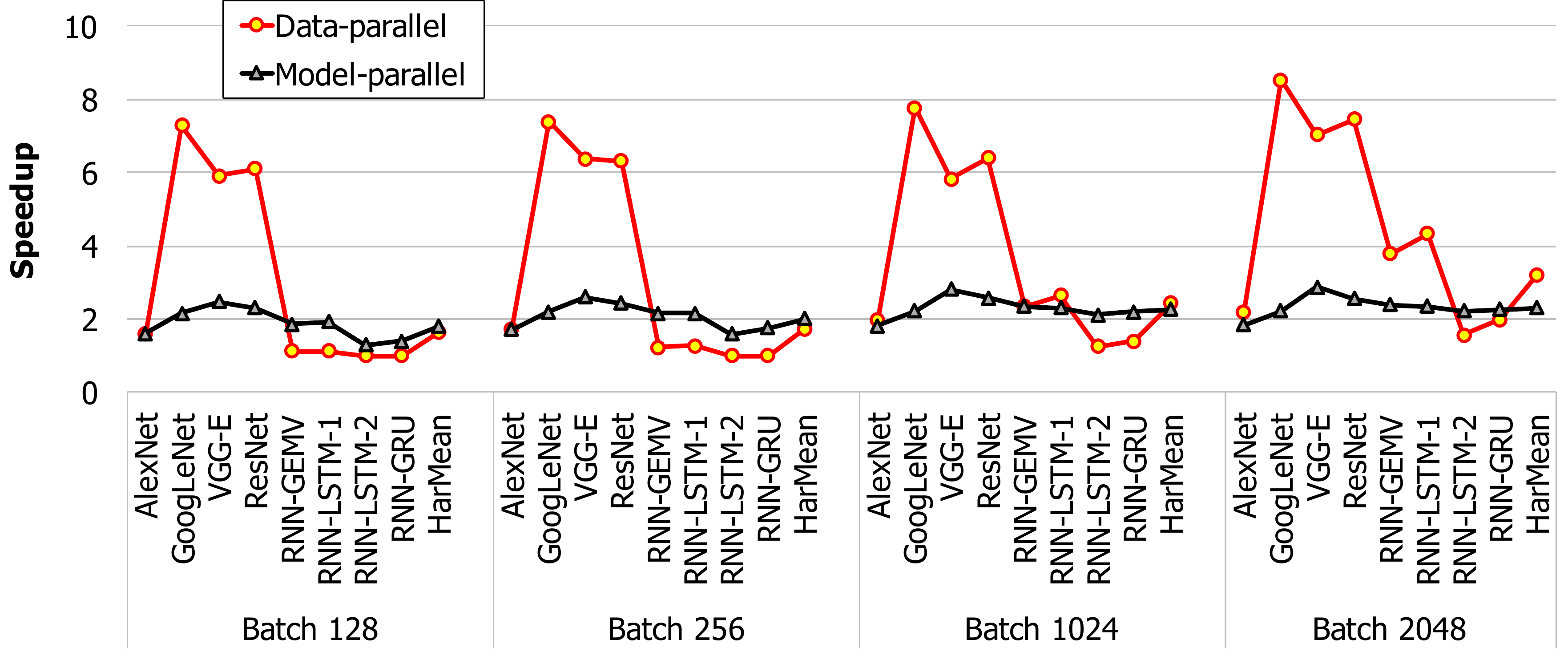}
\caption{
\mcdls performance sensitivity to input batch size.
}
\vspace{-1.2em}
\label{fig:sensitivity}
\end{figure}

{\bf Sensitivity.} \fig{fig:sensitivity} shows \mcdlsB's performance
sensitivity to the input batch size, demonstrating \mcdls's robustness (an
		average $2.17\times$ speedup over \dcdls across all batch sizes).  We also
studied \dcdls with the next generation PCIe (gen4) which doubles the PCIe link
bandwidth and improves \dcdls's memory virtualization performance. Such design
point improves \dcdls's performance by $38\%$, narrowing the performance gap
between \dcdls and \mcdls to $2.1\times$ (as opposed to $2.8\times$), but comes
at a cost of significant CPU memory bandwidth consumption (proportional to the
		increase in PCIe link bandwidth).  System designs with (1) a faster device-node
configuration such as TPUv2 and (2) scaled-up node configuration such as DGX-2 ($2$ PFLOPS and 
$2.4$ TB/s of device-side interconnect bandwidth\footnote{By provisioning even higher compute and
communication bandwidth than the baseline system, the benefits of \mcdls is even more pronounced as \dcdls becomes 
completely bottlenecked by memory virtualization.}) have also been explored which leads to \mcdls with
an average $3.2\times$ and $2.9\times$ speedup over \dcdls, respectively. 
Rhu et al.~\cite{rhu:2018:cdma}
proposed to leverage CNN activation sparsity to compress and reduce
local-device communication traffic to alleviate the PCIe bottleneck. This technique
provides an average $2.6\times$ reduction in PCIe traffic, narrowing the performance gap
between \dcdls and \mcdls to $2.3\times$ for the $4$ CNN applications we study.
Overall, \mcdls exhibited robustness across various sensitivity studies we
conducted (e.g., different chip configurations, input batch sizes, and etc.) as
it guarantees high-performance memory virtualization and inter-device
communication by design.

\subsection{Power Efficiency}
\label{sect:power}

\mcdls utilizes existing accelerators as-is, so the major power
overhead comes from the memory-nodes added to the ring network.
NVIDIA's DGX system (i.e., \dcdls) has a TDP of $3,200$ W, where the eight V100
GPUs consume $75\%$ of the system power (i.e., $300$ W $\times$ $8$).
\tab{tab:power} summarizes our estimation of a single memory-node's power
consumption using publicly available power measurements of DDR4
DIMMs~\cite{ddr4_measure} and Micron's DDR4 system power
calculator~\cite{micron_dram_power_estimator}. For power-limited environments,
	memory-nodes with $8$ GB RDIMM would be most appropriate which incurs an
	additional ($29$ $\times$ $8$) = $232$ W power consumption ($7\%$ increase
			over DGX-1V). For consumers more so focused on capacity expansion, the
	$128$ GB LRDIMM based memory-node would provide high value ($1.3$ TB of
			memory under $127$ W, highest GB/W). System-wide power consumption will increase by
	$31\%$ (i.e., $127$ $\times$ $8$ = $1,016$ W), but such option would
	drastically increase the pool of memory by $10.4$ TBs.  Microsoft's
	custom-built HGX-1~\cite{hgx_1}, a \texttt{4U} server chassis featuring $8$
	Pascal GPUs, can have a TDP up to $9,600$ W, so we believe the design overheads of
	\mcdls is reasonable given its unique value proposition.
	Overall, \mcdls achieves
	 ($2.8\times$/$1.31$)=$2.1\times$ to ($2.8\times$/$1.07$)=$2.6\times$ increase in performance per watt
	 while substantially enhancing the pool of memory exposed to the device-nodes.

\begin{table}[t!]
  \centering
  \caption{Memory-node power consumption (DDR4-2400). }
%\small
\footnotesize
%\scriptsize
%\vspace{-1em}
  \begin{tabular}{|c|c||c|c|}
    \hline
		\multicolumn{2}{|c||}{\textbf{Single DIMM}} & \multicolumn{2}{|c|}{\textbf{Memory-node}}  \\
		\hline
    \textbf{DDR4 modules} & \textbf{TDP (W)} & \textbf{TDP (W)} & \textbf{GB/W} \\
    \hline
    \hline
    $8$ GB RDIMM~\cite{rdimm_8gb}	& $2.9$  & $29$ & $2.8$ \\
    \hline
    $16$ GB RDIMM~\cite{rdimm_16gb}	& $6.6$ & $66$ & $2.4$\\
    \hline
    $32$ GB LRDIMM~\cite{lrdimm_32gb}	& $8.7$ & $87$ & $3.7$ \\
    \hline
    $64$ GB LRDIMM~\cite{lrdimm_64gb}	& $10.2$ & $102$ & $6.3$ \\
    \hline
    $128$ GB LRDIMM~\cite{lrdimm_128gb}	& $12.7$  & $127$ & $10.1$ \\
    \hline
  \end{tabular}
\vspace{-1.2em}
  \label{tab:power}
\end{table}

%------------------------------
\subsection{Scalability}
\label{sect:scalability}

Although the image classification problem~\cite{dnn_train} is gradually gaining
less traction from the DL algorithm community, there is still on-going research
in parallelizing and distributing CNN training to $1000$s of GPUs/TPUs to
reduce training time and achieve performance scalability.  Recent advances in
this domain of research~\cite{cnn_scaling_1,cnn_scaling_2,cnn_scaling_3,intel_sysml} employ
data-parallel training with extremely large batch sizes (e.g., $32$K in
		\cite{cnn_scaling_2}) to reduce the intra-/inter-node communication
overheads and achieve near perfect performance scaling. As the memory usage of
these existing CNN algorithms are optimized to fit within the physical GPU
memory constraints, training is done without any CPU-GPU data migration
involved. Using our simulation infrastructure, we observe similar (perfect) performance
scalability with \dcdls when memory virtualization is disabled (i.e., close to
		$4\times$ and $8\times$ reduction in training time when the $4$ CNN
		applications in \tab{tab:benchmarks} are data-parallelized across $4$/$8$
		GPUs). However, when memory virtualization is enabled and the feature
maps are migrated in/out of local-remote memory, the performance improvements
achieved with $4$/$8$ GPUs under \dcdls is only $1.3\times$/$2.7\times$ because
of the host-device communication bottleneck\footnote{Although host-device data migration for these CNN workloads is arguably 
		unnecessary, following prior work~\cite{rhu:2016:vdnn,park:2018:ppopp,lms_sysml,nikolai:gtc:2017},
		we use existing workloads to study performance scalability as
		there are no publicly available DNN algorithms that exceed
the memory capacity limits of current systems (i.e., you cannot train a DNN algorithm
unless its memory requirement fits within the physical memory size limits, the chicken-and-egg problem).
}. Performance scalability is regained using \mcdls thanks to its ability to perfectly hide data migration
overhead (\fig{fig:latency_breakdown}).

%------------------------------
\subsection{User Productivity}
\label{sect:productivity}

As state-of-the-art CNN algorithms for \emph{image}
classification~\cite{resnet,densenet} reach super-human performance, the
DL research community has shifted towards
more challenging tasks such as \emph{video} understanding (e.g., video
		classification and
		captioning~\cite{seq2seq_video_to_text,video_caption_book,yu:2016:cvpr},
		video question and
		answering~\cite{read_write_mem_network,movieqa,zeng:2017:aaai}).  Given an
input video stream, the goal is to capture the context of the
scenes, objects, and activities and be able to express how these
relate to each other in a complete sentence.  State-of-the-art video
understanding algorithms are commonly implemented as a mixture of CNNs, LSTMs,
							and memory networks~\cite{neural_turing_machine,dynamic_ntm}, but
							training these algorithms end-to-end under current HPC systems
							becomes practically impossible because of the memory capacity
							bottleneck. DL practitioners are therefore forced to compromise
							their learning algorithm (e.g., freezing subset of the algorithm
									without end-to-end training, reducing the number of input
									video frames and recurrent timesteps per training iteration,
									cropping video frame sizes, \ldots) so that the overall
							memory footprint fits within the physical GPU memory. With the
							advent of large-scale video training datasets such as
							YouTube-$8$M~\cite{youtube_8m}, providing sufficient amount of
							memory that enhances user productivity will become
							vital.  Aside from being able to train DNNs that are
							deeper and larger, \mcdls can open up a wider
							range of complex learning algorithms (e.g., end-to-end training of
									aforementioned video-to-text algorithms employing larger
									CNNs/LSTMs) that are currently impossible to train due to 
							memory capacity limits, propelling continued innovation in this active
							research space.

\begin{figure}[t!] \centering
\includegraphics[width=0.47\textwidth]{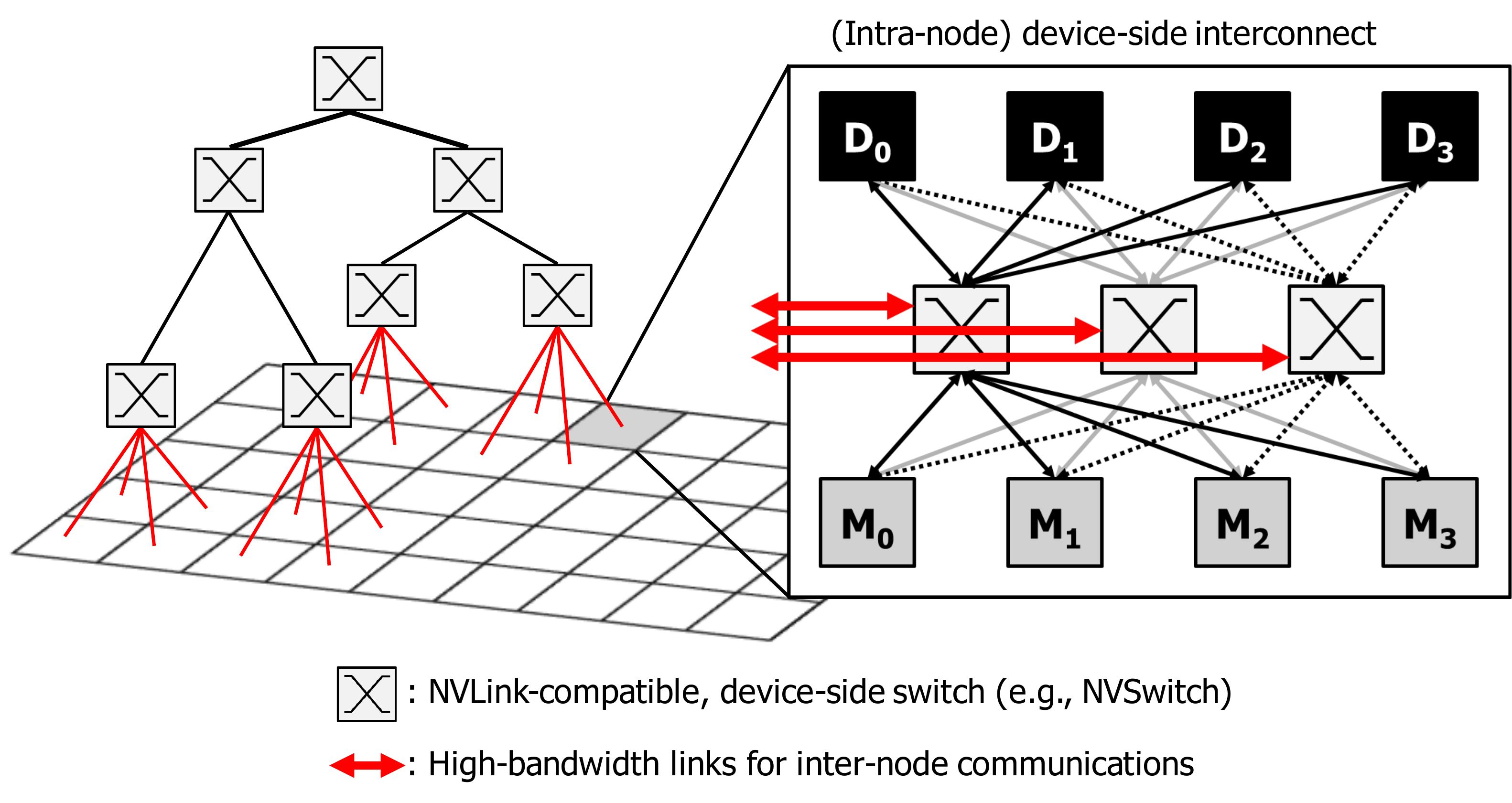} \caption{
Scale-out, datacenter-level device-side interconnect plane. Figure assumes that
a given system node houses $8$ nodes and each device-/memory-node is
provided with \texttt{N}=$3$ high-bandwidth links. The
device-side switch allows each of the $8$ nodes to communicate with any of the
	other nodes inside the system node, enabling it to be casted into any interconnection 
topology (e.g., the ring-based \mcdls interconnect,
			\fig{fig:mcdla_design_space}(c)).  } \vspace{-1.5em}
	\label{fig:mcdla_scaleout} \end{figure}

\section{Future Research Direction}
\label{sect:future}

To the best of our knowledge,
	our work is the first in the literature that highlights the growing
	significance of device-side interconnects in training DL algorithms
	across multiple devices. 
Due to space limitations and the wide design space, 
	this paper focused on \emph{intra}-node system architectural issues,
	assuming \emph{inter}-node communication is handled using MPI over Ethernet/InfiniBand.  
	NVIDIA recently announced the NVSwitch~\cite{nvswitch}
	technology which is an NVLINK-compatible switch, enabling system
	vendors opportunities to scale-up/out device-side
	interconnection networks, for instance (1) incorporating a larger number of
	GPUs within a system node~\cite{dgx_2} or (2) tightly integrating thousands of
	GPUs across hundreds of system nodes  (\fig{fig:mcdla_scaleout}), similar to
	Microsoft's BrainWave~\cite{brainwave}.  The introduction of
	these device-side switching technologies for accelerators that enable
	scale-out device-side interconnects emphasizes the
	importance of device-side interconnection networks moving forward and opens up
	interesting research opportunities. 
	Exploring our memory-node architecture as part of these
	scale-out device-side interconnects with $100$s of device-nodes and
	memory-nodes across a distributed network is part of our next future work.

\section{Related Work}
\label{sect:related}

Memory disaggregation~\cite{disagg_mem_1,disagg_mem_2} expands the CPU memory
hierarchy to include a remote level provided by a separate memory blade
connected over PCIe, which helps increase the pool of CPU accessible memory. 
Kim et al.~\cite{kim:2013:mcn,kim:2014:mn_gpu} proposed to
interconnect multiple CPUs/GPUs by leveraging the packet routing capabilities
of HMCs~\cite{hmc}, effectively composing a memory network that
provides flexible processor bandwidth utilization. 
The scope of \cite{kim:2013:mcn,kim:2014:mn_gpu} is significantly different than what our work
focuses on, but more importantly, our proposal is not tied with a
particular memory technology whereas \cite{kim:2013:mcn,kim:2014:mn_gpu}
assumes a $3$D stacked memory with routing capabilities embedded inside the
logic layers.
Slowdown on Moore's law and Denard scaling have driven researchers to pursue ``chiplet'' 
based processor designs~\cite{amd_chiplet:micro:2016,mcm_gpu,amd_chiplet:isca:2018}, where a large
SoC is decomposed into multiple smaller (but higher yield) chiplets and are
re-assembled as a single package. One can envision combining the concept of chiplet-based 
GPUs with the notion of memory networks~\cite{kim:2013:mcn} as means to tightly integrate
GPUs and HMCs within a package for memory capacity expansion. However, the maximum number of GPUs 
as well as HMCs that can be integrated inside a single package is bounded by various technology
constraints, load distribution, and ease of programmability (e.g., recent MCM-GPU assumes only up-to $4$ GPUs integrated within a single package).
The focus of \mcdls is on efficient parallelization and workload partitioning in a system-level context
as opposed to these prior chiplet-context studies focusing on package-level or board-level integrations.
A large body of prior work has explored the design of  
a single accelerator device architecture for deep learning
inference~\cite{dadiannao,eyeriss,diannao,shidiannao,pudiannao,du:2015:micro,minerva,dnn_pim_reram,cambricon,neurocube,tabla,dnnweaver,intel:2017:fpl,gao:2017:tetris,intel:2018:fpga}
with an increased interest on leveraging DNN sparsity for further energy-efficiency improvements~\cite{scnn,song:2015:eie,cnvlutin,cambriconx,stripes,bitpragmatic,intel:2017:icassp,intel:2017:fpga,whatmough:2017:isscc,whatmough:2017:hotchips,bittactical}.
Park et al.~\cite{cosmic} proposed a scale-out acceleration
platform for training machine learning algorithms using an FPGA-based $16$-node distributed system.
These prior studies are orthogonal to our \mcdls proposal and can be adopted  
further for additional enhancements.

\section{Conclusion}
\label{sect:conclusion}

As the models and datasets to train DL models scale,
system vendors are employing a custom
device-side interconnection network for fast communication and 
synchronization across accelerator devices. 
This paper is the first to describe the growing significance of
device-side interconnects for training scaled up DL algorithms
and highlights the importance of
balancing inter-device communication and fast memory virtualization. We make
a case for a memory-centric DL system and presented  
a scalable, programmable, and energy-efficient HPC platform for DL training,
which provides
an average $2.8\times$ speedup over \dcdls while drastically
expanding the pool of memory accessible to accelerators to $10$s of TBs.

\section*{Acknowledgment}
This work was supported by Samsung Research Funding Center of Samsung Electronics under Project Number SRFCTB1703-03.

%\bibliographystyle{ieeetr}
%\bibliography{ref}

%\begin{thebibliography}{1}

%\end{thebibliography}

\end{document}